\documentclass[12pt]{article}\usepackage[hyperfootnotes=false]{hyperref}
\usepackage{epsfig}
\usepackage{float}
\usepackage{empheq}
\usepackage{bbold}

\usepackage[utf8]{inputenc}
\usepackage{amsmath}

\usepackage{caption}

\usepackage{amsmath}
\usepackage{amssymb}
\usepackage{graphicx}
\newlength{\abstractwidth}
\renewcommand{\thefootnote}{\fnsymbol{footnote}}
\renewcommand{\thanks}[1]{\footnote{#1}} 
\newcommand{\starttext}{
\setcounter{footnote}{0}
\renewcommand{\thefootnote}{\arabic{footnote}}}

\newcommand{\be}{\begin{equation}}
\newcommand{\bea}{\begin{eqnarray}}
\newcommand{\eea}{\end{eqnarray}}
\newcommand{\beq}{\begin{equation}}
\newcommand{\ee}{\end{equation}}

\newcommand*\widefbox[1]{\fbox{\hspace{2em}#1\hspace{2em}}}

\def\eq{&=&}

\def\la{\langle}
\def\ra{\rangle}
\def\simleq{\; \raise0.3ex\hbox{$<$\kern-0.75em
\raise-1.1ex\hbox{$\sim$}}\; }
\def\simgeq{\; \raise0.3ex\hbox{$>$\kern-0.75em
\raise-1.1ex\hbox{$\sim$}}\; }

\def\bi{\begin{itemize}}
\def\ei{\end{itemize}}
\def\S{Schwarzschild}

\def\dof{degrees of freedom }

\def\CG{{\cal{G}}}
\def\CH{{\cal{H}}}

\def\CJ{{\cal{J}}}
\def\CK{{\cal{K}}}

\def\CO{{\cal{O}}}
\def\CP{{\cal{P}}}

\def\CT{{\cal{T}}}

\def\t{\tau}

\def\Tr{\rm Tr \it}
\def\bsub{ \begin{subequations}
\begin{empheq}[box=\widefbox]{align}  }
\def\esub{ \end{empheq}
\end{subequations}}

\def\1{\(  \mathbb{1} \)}

 \def\lf{\left(}
    \def\rg{\right)}
    \def\b{\bf{b}}

  \def\bn{\bigskip \noindent}

\def\a{\bf{a}}                
 \def\p{\bf{p}\it}

 \def\bm{\begin{bmatrix}}
 \def\em{\end{bmatrix}}

  \def\p{{\bf{p}}}
    \def\a{{\bf{a}}}
    \def\dS{de Sitter space \ }

\makeatletter
\g@addto@macro\normalsize{%
  \setlength\abovedisplayskip{10pt}
  \setlength\belowdisplayskip{20pt}
  \setlength\abovedisplayshortskip{10pt}
  \setlength\belowdisplayshortskip{20pt}
}
\makeatother

\usepackage{color}

\begin{document}


  
\begin{titlepage}

\rightline{}
\bigskip
\bigskip\bigskip\bigskip\bigskip
\bigskip

\centerline{\Large \bf {De Sitter Holography:}} 

\bn

\centerline{\Large \bf {Fluctuations,  Anomalous Symmetry,  and Wormholes }}

\bigskip
\begin{center}
\bf      Leonard Susskind  \rm

\bigskip
Stanford Institute for Theoretical Physics and Department of Physics, \\
Stanford University,
Stanford, CA 94305-4060, USA \\

and

Google, Mountain View, CA

\end{center}

\bn

\begin{abstract}

The Goheer-Kleban-Susskind no-go theorem says that the  symmetry of de Sitter space is incompatible with finite entropy. The meaning and consequences of the theorem are discussed in the light of recent developments in holography and gravitational path integrals. The relation between the GKS theorem, 
Boltzmann fluctuations,   wormholes, and exponentially suppressed non-perturbative phenomena suggests:  the classical symmetry between different static patches is broken; and  that eternal de Sitter space---if it exists at all---is an ensemble average.  

\end{abstract}

\end{titlepage}

\starttext \baselineskip=17.63pt \setcounter{footnote}{0}

\Large

\tableofcontents


\section{Introduction}

\bn
\it All phenomena in a region of space can be described by a set of degrees of freedom localized on the boundary of that region, with no more than one degree of freedom per Planck area. \rm
 
 \bn
The Holographic Principle has been the driving force behind many of the advances in quantum gravity over the last twenty five years, but
up to now the only precise examples have been  cosmologies, which  like anti de Sitter space,    have asymptotically cold\footnote{By cold I mean non-fluctuating. By warm I mean the opposite.}  time-like causal boundaries. The reliance on the existence of such   boundaries is troubling because the space we live in seems not to have one. Instead it has a horizon and a space-like warm boundary. If the HP is to apply to the real world then we need to generalize it.

So, does the Holographic Principle apply to cosmologies like de Sitter space? If yes, then what are the rules?  I don't know for sure, and this paper will not conclusively answer the question, but I will try to lay out some tentative principles. 

Two things that will not be found here are specific models and applications to phenomenology.

\subsection{An Obstruction? Or Not}
Some time ago  Goheer, Kleban, and I (GKS) \cite{Goheer:2002vf}  proved a theorem that  it is impossible for a quantum system to satisfy the symmetries of classical de Sitter space if the entropy is finite\footnote{The framework for \cite{Goheer:2002vf} was the same as for this paper; namely the assumption that a static patch of de Sitter space has  a holographic description based on conventional Hamiltonian quantum mechanics.  }. At the time, I interpreted this as a no-go theorem for absolutely stable (eternal) de Sitter space, but  recent developments in quantum gravity  suggest that a different interpretation might be possible. The idea is   that a de Sitter vacuum  might be eternal but   the symmetries  only approximate, being violated by exponentially small non-perturbative effects. The mechanisms are very similar to ones that have recently been uncovered in the SYK system and its gravitational dual.

\subsection{Eternal de Sitter Space} \label{EDS}

\begin{figure}[H]
\begin{center}
\includegraphics[scale=.4]{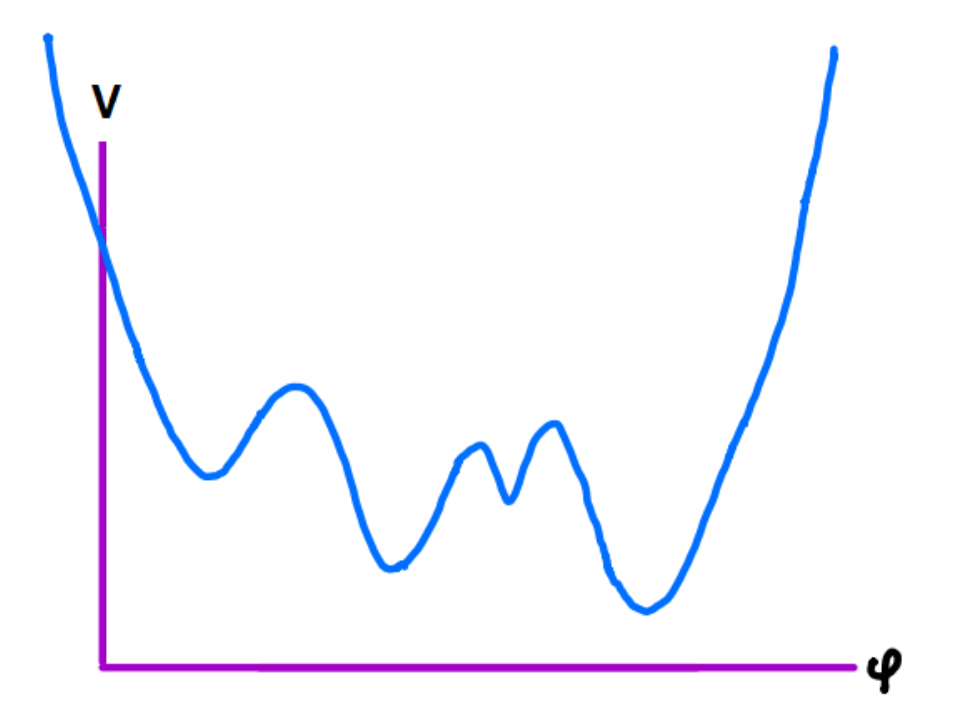}
\caption{A schematic landscape for eternally stable de Sitter space. The universe spends most of its time in the lowest minimum with rare fluctuations to higher points.}
\label{landscape}
\end{center}
\end{figure}

By  eternal \dS  I mean  a cosmology that is trapped in a state of finite entropy and cannot escape through reheating or tunneling to a larger ``terminal vacuum" 
 \cite{Bousso:2006ge}\cite{Harlow:2011az}\cite{Susskind:2012pp}\cite{Susskind:2012yv}\cite{Susskind:2012xf}.
Eternal \dS might arise from a landscape in which the scalar fields have strictly positive potential, 
greater than some finite positive gap, in which there is one or more minima of $V.$  Figure \ref{landscape} illustrates this kind of potential. I don't know if a landscape with these properties can  exist in a real theory of quantum gravity but let's assume that it can, and see where it takes us.

\bn
Classically the different minima lead to different stable  de Sitter geometries, but in quantum mechanics tunneling allows transitions between the minima. This leads to a single thermal equilibrium  state (of a static patch) which mostly sits in the lowest minimum. But occasional  Boltzmann fluctuations allow transitions to higher minima, followed by tunneling back to the
 lowest minimum. The rates for such  fluctuations are of order $\exp{(-S_0)}$ where $S_0$ is
the nominal  entropy of the de Sitter space at the lowest minimium\footnote{The notation $\exp$ will be used to indicate a general
     exponential scaling. Thus $e^S,$  $e^{S/3,}$ and $e^{2S}$ are all $\exp{(S)}$. }. Other less extreme fluctuations can take place; for example the horizon of a static patch  may spontaneously emit an object such as a black hole. These freak fluctuations are the only things that happen in the closed quantum world of a de Sitter static patch  \cite{Dyson:2002nt}\cite{Dyson:2002pf}, but  thermal correlations 
contain a wealth of information about non-equilibrium dynamics.

Eternal \dS is of course eternal, both to the future and to the past\footnote{From the viewpoint of a static patch.}.  Among the possible Boltzmann fluctuations that can take place over an enormous expanse of time are transitions to what we  ordinarily think  of as the initial conditions of our universe. I have in mind a fluctuation to to an inflationary point on the landscape, which after some inflation eventually evolves to a standard $\Lambda$-CDM universe. That of course would  not be the end of the story. After a very long period of $\Lambda$-dominance a fluctuation will occur to another minimum and the whole process will repeat.

While this may be possible,
a theory based on Boltzmann fluctuations is a very implausible framework for cosmology. In order to escape the eternal cycle of recurrences, near recurrences,   partial recurrences, and freak histories described in \cite{Dyson:2002pf} it seems necessary to have a landscape that includes terminal vacua  \cite{Bousso:2006ge}\cite{Harlow:2011az}.  A phenomenon called ``fractal flow"  \cite{Susskind:2012pp}\cite{Susskind:2012yv}\cite{Susskind:2012xf}   can then lead to a much more plausible cosmology. 

Nevertheless it is interesting to explore the consequences of the Holographic Principle for eternal de Sitter space even if it eventually leads to the conclusion that eternal de Sitter space is inconsistent. 

\section{Static Patch Holography}
There are a number of different approaches to de Sitter space that might loosely be called holographic. I will stick to the original meaning of the term: a description localized on the boundary of a spatial region in terms of a  quantum system without gravity. Specifically  we will focus on  static patches and their boundaries---cosmic  horizons. I will not  speculate on the details of the quantum system other than to say it should be fairly standard;
 for example it might be described as a collection of qubits, or some form of matrix quantum mechanics with a Hermitian Hamiltonian. The bulk space-time and its geometry   emerge from the holographic degrees of freedom.

\subsection{The Semiclassical Limit}
The classical limit of de Sitter space is described by the metric,
\bea 
ds^2 \eq -f(r)dt^2 +f(r)^{-1}dr^2 +r^2 d\Omega_2^2 \cr \cr
f(r) \eq \lf 1- \frac{r^2}{R^2}  \rg
\label{SPmetric}
\eea
The  length-scale $R$ is the radius of curvature, inverse to the Hubble parameter.  The cosmic horizon at $r=R$ is the place where $f(r)=0.$

The semiclassical limit refers to the theory of small perturbations about the classical geometry, which can be described in powers of $\hbar.$ Although it may be sufficient for many purposes the semiclassical theory is incomplete.
The full quantum theory will have  non-perturbative effects of magnitude $$\exp{(-\frac{a^2}{G\hbar})},$$ where $a$ is some characteristic length scale and $G$ is Newton's constant.  If $a \sim R$ then the nonperturbative effects are order $$\exp{(-S_0)},$$  $S_0$ being the de Sitter entropy. In the semiclassical limit  only the zeroth  order term in $\exp{(-S_0)}$ is retained. 

De Sitter space is in some ways similar to a black hole. Both have horizons, an entropy proportional to the horizon area, and a temperature. Both have a semiclassical limit and additional non-perturbative effects.
For any real black hole the nonperturbative effects are exponentially small (in the entropy),  but they play a crucial role  in establishing the consistency between quantum mechanics and gravity. It seems reasonable that  the same would true for de Sitter space. Sections \ref{Sec: toy} through \ref{dS} are about these  nonperturbative effects but for now we focus on the semiclassical theory.

\begin{figure}[H]
\begin{center}
\includegraphics[scale=.35]{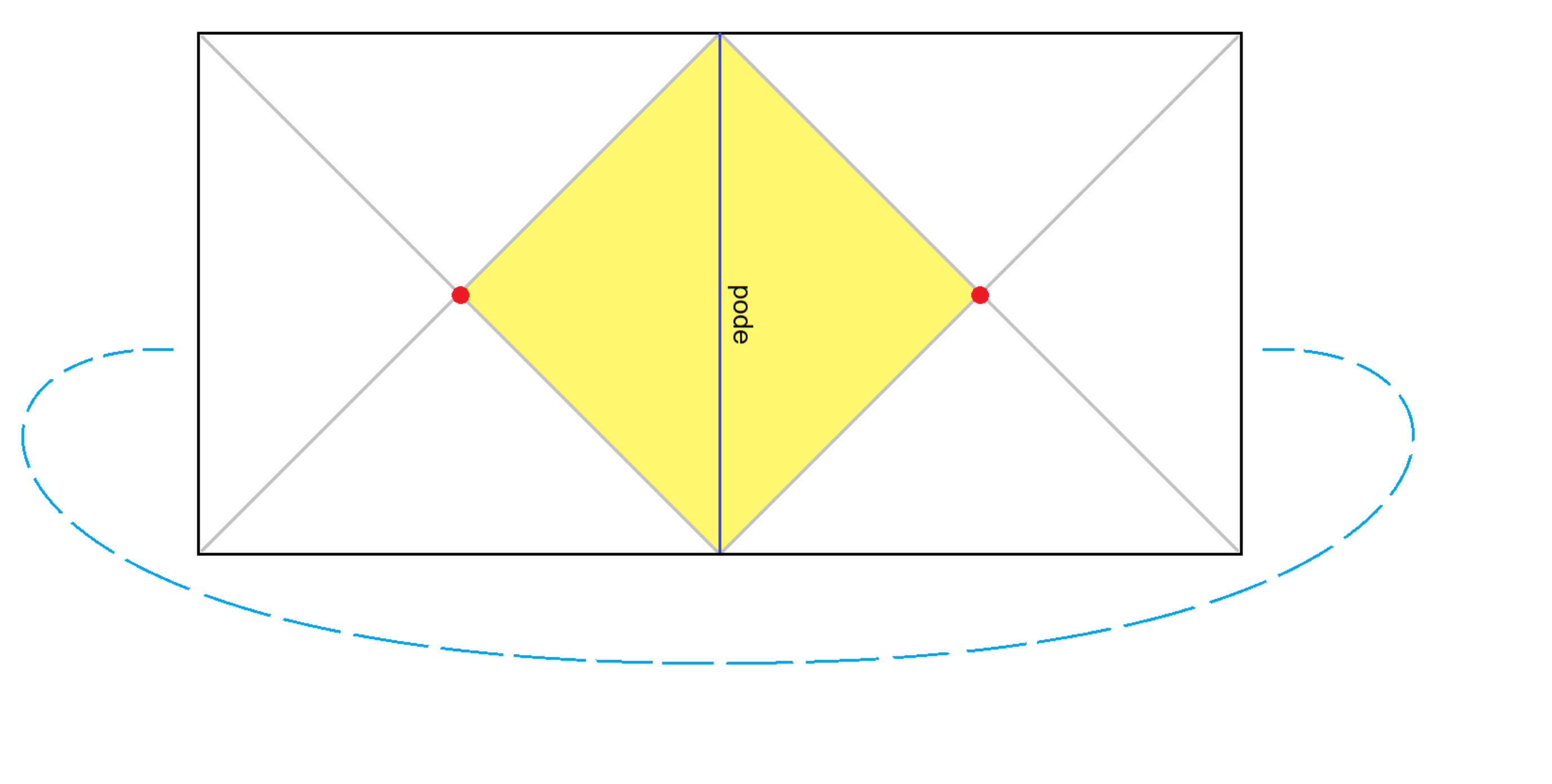}
\caption{Conformal diagram for 2-D de Sitter space and the static patch defined by a past and future pair of asymptotic points. The static patch (yellow) is the intersection of the causal future of the past point and the causal past of the future point. The intersection of the two light cones shown as red dots defines the bifurcate horizon. The dashed blue curve indicates identification of the left and right edges.}
\label{2Dpenrose}
\end{center}
\end{figure}

To illustrate the static patch consider  the example of $(1+1)$-dimensional de Sitter space. The conformal  diagram is a rectangle twice as wide as it is high, and periodically identified\footnote{Strictly speaking a true Penrose diagram would only contain a single square with each point on the diagram representing a zero-sphere, i.e.,  two points. } as illustrated in fig \ref{2Dpenrose}. \\
A static patch is defined by picking a pair of points\footnote{A ``point" on the asymptotic boundary does not literally mean a point but rather a region of proper size no bigger than $R$.  This is familiar in AdS/CFT where a point on the boundary has $N^2$ degrees of freedom and represents a region of $AdS$ size \cite{Susskind:1998dq}. }, one on the asymptotic past and one on the asymptotic future. The static patch is the intersection of the causal future of the past point and the causal past of the future point. The geodesic connecting the asymptotic points will be referred to as the world-line of the ``pode."

Observers who spend their entire existence in the static patch will see their world bounded by the horizon although the full geometry has no boundary. Our central hypothesis is that  everything that goes on in the  static patch can be described  by a unitary  holographic system with the degrees of freedom  located at the  stretched horizon (see section \ref{where is}). The holographic quantum mechanics, which includes a Hilbert space and  a Hermitian Hamiltonian, allows us to define certain thermal properties of the static patch including a  density matrix, a  temperature $T=1/\beta$, and an entropy $S_0$.

\bea 
\rho \eq \frac{e^{-\beta H}}{Z} \cr \cr
\beta \eq \frac{1}{T} =2\pi R \cr \cr
S_0\eq \Tr   \ \rho \log{\rho} =\frac{\pi R^2}{G}
\label{dSprops}
\eea

\begin{figure}[H]
\begin{center}
\includegraphics[scale=.5]{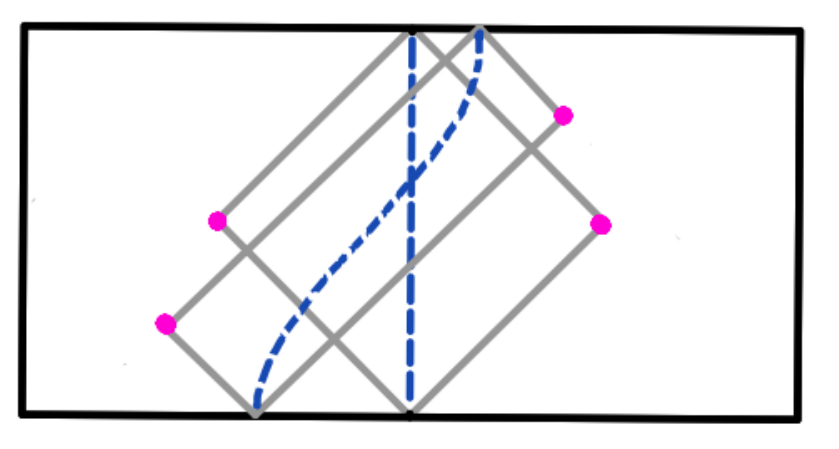}
\caption{Two static patches in the same dS.}
\label{transformation}
\end{center}
\end{figure}

In addition to the quantum mechanics of a single static patch, we also require transformation laws between static patches. For example in fig \ref{transformation} we see two static patches of $dS_2$. 
\bn
A theory of \dS should have transformation rules  relating conditions in different static patches. In classical GR these transformations are symmetries that express the identical nature of the  patches. These symmetries relating static patches, and the possibility of representing them in a holographic theory are the main subject of this paper.

\begin{figure}[H]
\begin{center}
\includegraphics[scale=.50]{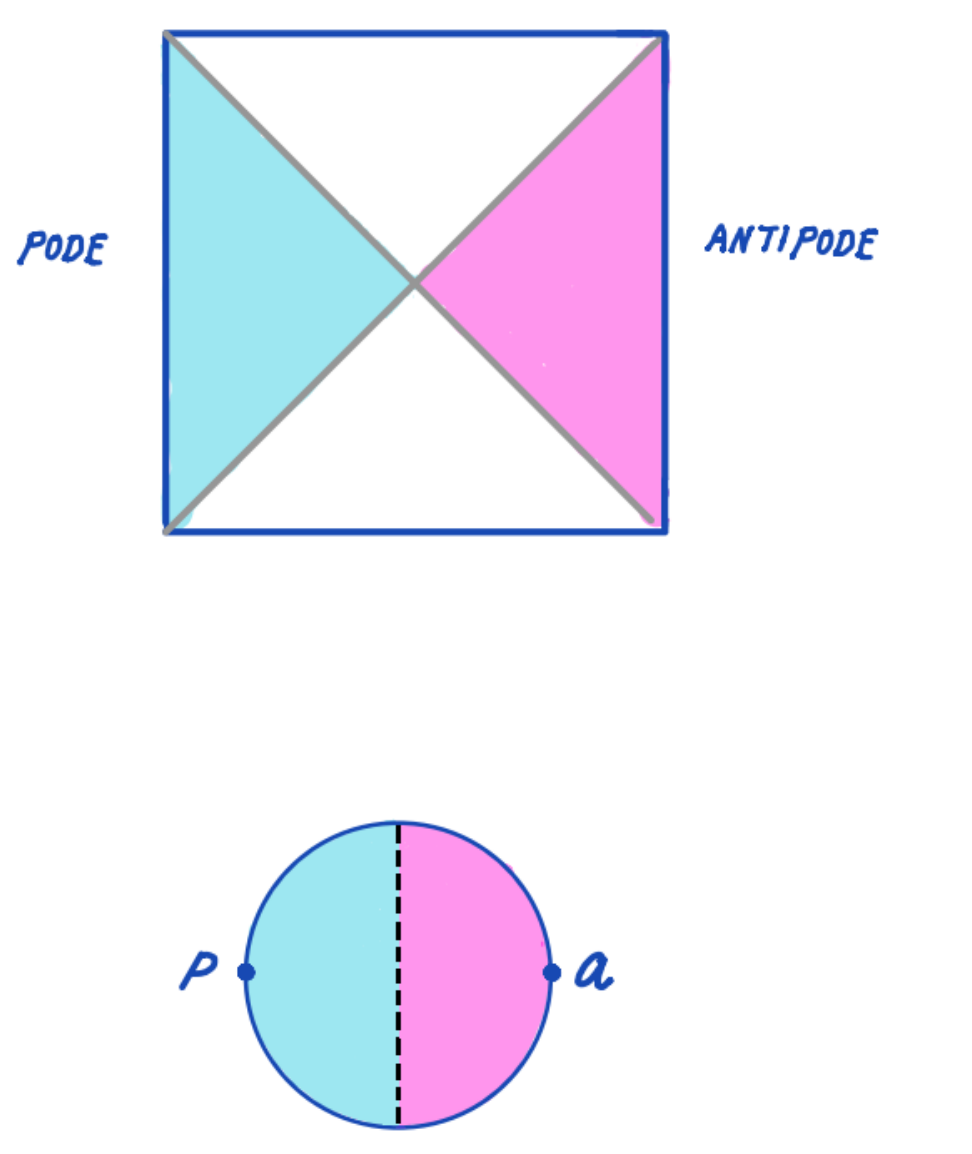}
\caption{Upper panel: Penrose diagram for higher dimensional de Sitter space. Static patches come in pairs and the center of these patches are referred to as the pode and the antipode.
Lower panel: the geometry of the $t=0$ slice of \dS is a sphere with the pode at one pole and the antipode at the other. The dashed surface midway between the pode and antipode is the bifurcate horizon.}
\label{penrose}
\end{center}
\end{figure}

The Penrose diagram for general dimensional de Sitter space is shown in the top panel of fig \ref{penrose}. 

\bn
The diagram shows that static patches come in matched pairs---blue and pink in the diagram. We will refer to the points at the centers of these static patches as the pode and the antipode.

\subsection{Where is the Hologram?}\label{where is}

The penrose diagram of fig \ref{penrose} looks a lot like  the diagram for the two-sided AdS eternal black hole \cite{Maldacena:2001kr}. From all that we know about such black holes this suggests (and we will assume) that the podal and antipodal \dof \ are uncoupled, but  entangled in a thermofield-double state. However the geometries of the two-sided black hole and de Sitter space 
 are very different. In the lower panel of fig \ref{penrose}
we see a time-symmetric slice through the \dS geometry at  $t=0.$ The geometry of the slice  is a sphere.  By contrast the corresponding slice through the eternal black hole would be a wormhole connecting two infinite asymptotic boundaries. 
The spatial slice of \dS has no boundary, the pode and antipode being points at which the geometry is smooth. 

\begin{figure}[H]
\begin{center}
\includegraphics[scale=.5]{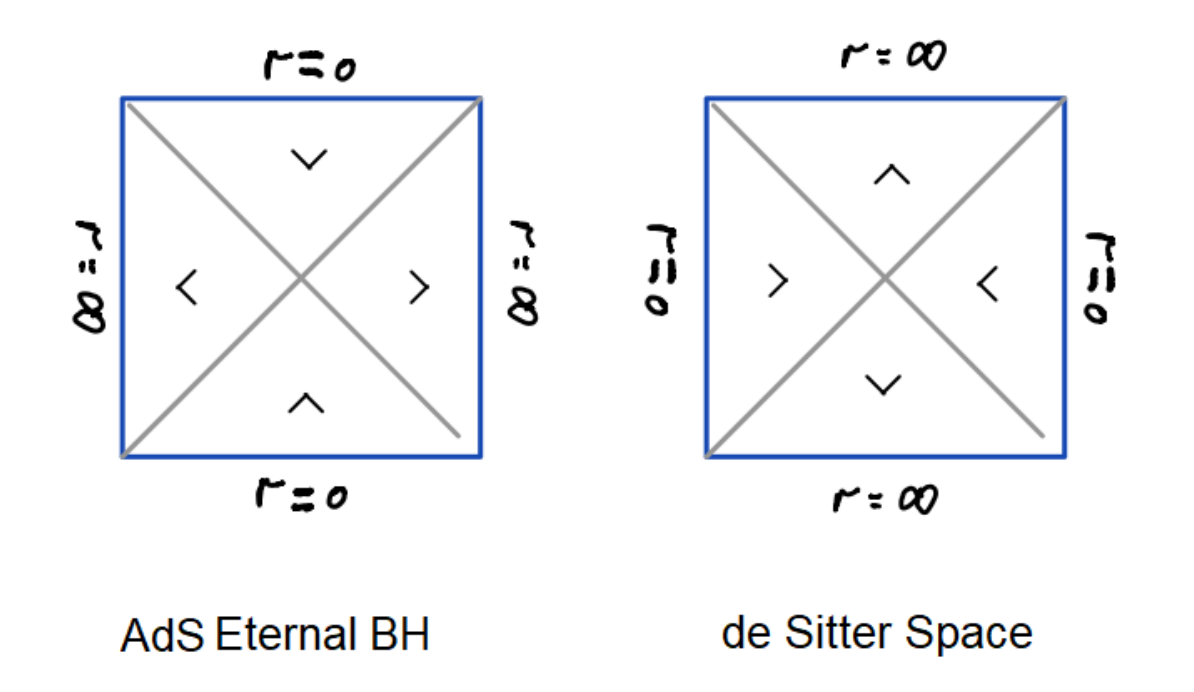}
\caption{Penrose diagrams supplemented with Bousso wedges for the AdS eternal black hole and for \dS.}
\label{bousso}
\end{center}
\end{figure}

\begin{figure}[H]
\begin{center}
\includegraphics[scale=.34]{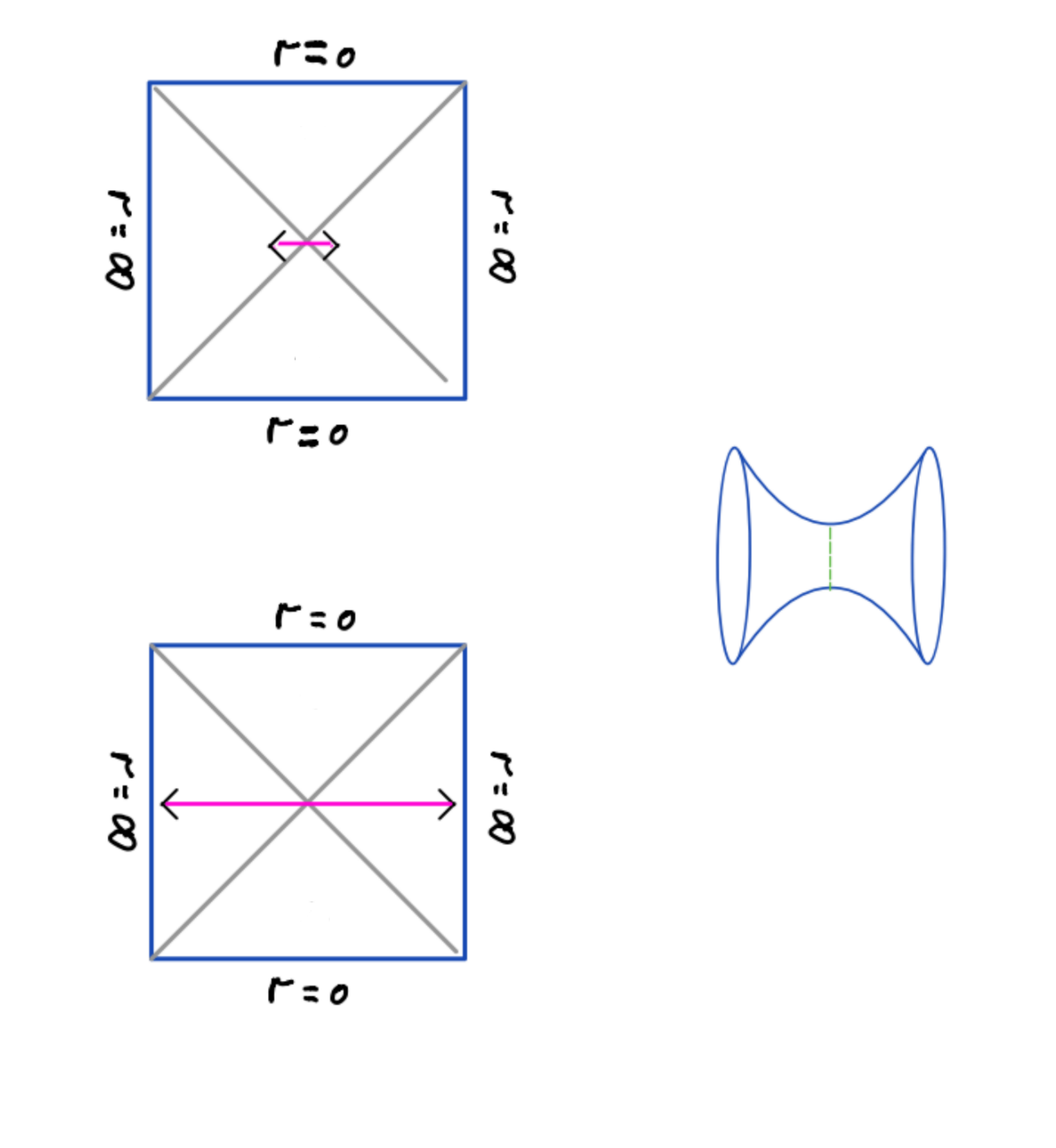}
\caption{The AdS eternal black hole case: the upper panel introduces  screens close to the horizon. According to the light-sheet entropy bound the number of \dof \  on the screens is sufficient to represent the states of the black hole, but not sufficient to represent \dof \ in the space far from the horizons. In the lower panel the screens are placed out near the AdS boundaries. In this case the \dof  \ on screens is sufficient to encode the entire space.}
\label{boussoads}
\end{center}
\end{figure}

Instead of being located at the boundary as in AdS the holographic \dof \ of the static patch are located at the (stretched)  horizon.
To see this, consider Bousso's generalization \cite{Bousso:1999xy}  of the Penrose diagrams  for the AdS eternal black hole, and for de Sitter space\footnote{Bousso supplements a Penrose diagram with a system of wedge-like symbols showing the direction in which light sheets focus and de-focus.}  These are shown  in fig \ref{bousso}.
The question is: Where should we locate the holographic screens  (tips of the wedges) so that the maximum entropy of the spatial region described by the hologram is sufficient to encode everything in the geometry?
In fig \ref{boussoads} the diagrams are shown for AdS in which we place the screens near the horizon in the first case and near the boundary in the second case. 

\bn
From the light-sheet  entropy bounds of  \cite{Fischler:1998st}\cite{Bousso:1999xy} one sees that in the first case the maximum entropy on the pink spatial  region is just a tiny bit larger than the black hole entropy. Placing holographic degrees of freedom at these locations would allow enough degrees of freedom to describe the black hole, but not enough to describe phenomena in the bulk far from the horizon.

In the second case, where the wedges are near the boundary, the maximum entropy grows as the screens are moved outward. This is the well-known reason that the holographic degrees of freedom of AdS  are located at the boundary.

\begin{figure}[H]
\begin{center}
\includegraphics[scale=.4]{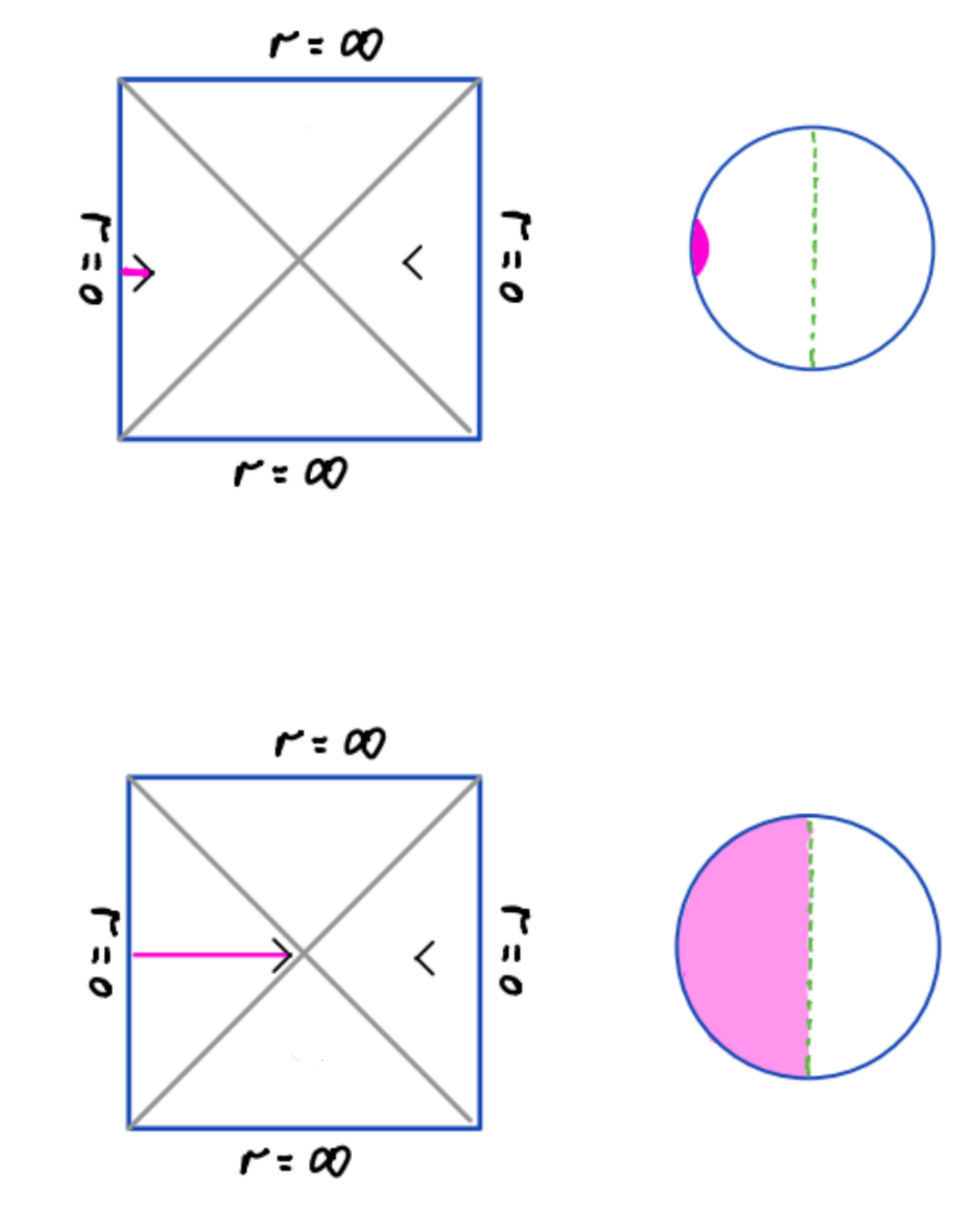}
\caption{Penrose-Bousso diagrams for \dS. In the upper panel the screen is near the pode and the screen has almost no \dof. In the lower panel the screen is near the horizon and the number of \dof \ is sufficient to encode the entire static patch.}
\label{boussods}
\end{center}
\end{figure}

Next  consider de Sitter space. In figure \ref{boussods} two choices for the locations of holographic screens are shown. In the upper panel the screen is shown near the pode.
\bn
The maximum entropy is very small when the screen is near the pode. By contrast, in the lower panel the screen is shown close to the horizon.The maximum entropy on the pink slice in this case is large enough to describe the entire static patch.That's the argument  for locating    the holographic \dof \ at the horizon.

\begin{figure}[H]
\begin{center}
\includegraphics[scale=.3]{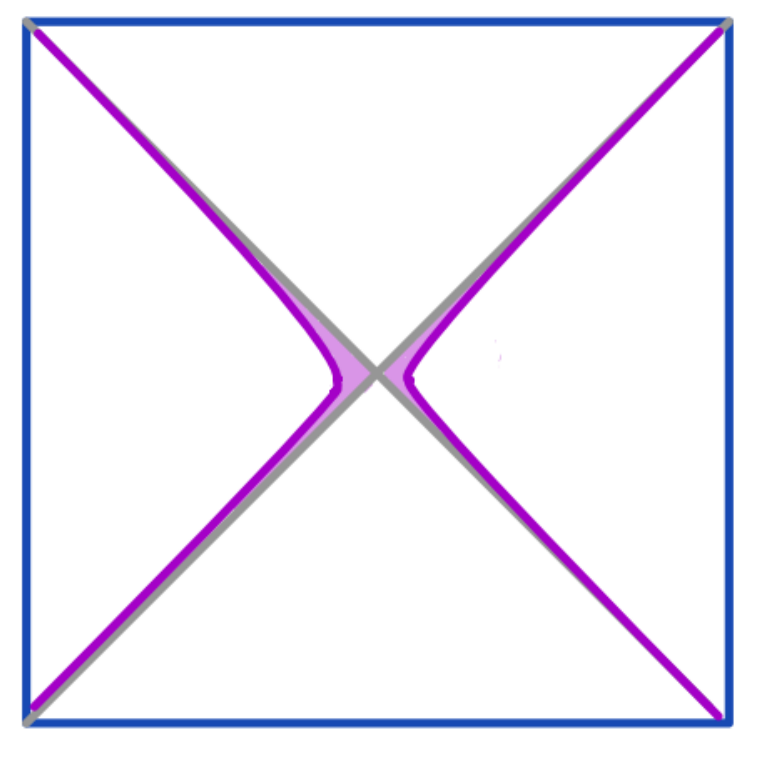}
\caption{The stretched horizons of the pode and antipode patches.}
\label{stretch}
\end{center}
\end{figure}

The Penrose diagram suggests that there are two sets of degrees of freedom---one for the pode and one for the antipode---located on the stretched horizons of each side, and entangled in a thermofield-double state (see fig \ref{stretch}).

\subsection{Symmetry of  De Sitter Space}\label{sym of dS}

Let's suppose that an oracle  handed us what he claimed is a holographic dual of de Sitter space. How would we know if it really describes \dS or something else---perhaps a black hole? The answer is symmetry. There are many static patches in \dS and the symmetry of the space transforms one to another. The most conclusive test  would be to show that the model satisfies the \dS symmetry. In figure \ref{transformation} we see that  one static patch can probe the region behind the horizon of another. Since the original pode-patch is described by  unitary evolution, establishing that the symmetry is satisfied would  tell us that all  static patches are described by unitary evolution. To put it another way, 
testing the de Sitter symmetry is also testing for the existence of  space-time   behind the horizon.  Without testing the symmetry there is no compelling reason to think that a given quantum system represents dS.

The symmetry of clasical $d$-dimensional de Sitter space is $O(d,1),$ a non-compact version of the orthogonal group $O(d+1).$ 
$O(d,1)$  has $$\frac{d(d+1)}{2}$$
generators. Of these $$\frac{(d-2)(d-1)}{2} +1$$ generate transformations that keep  the pode, antipode, and the  horizon fixed. I'll refer to these as the ``easy" generators. They include rotations $R$  about the pode-antipode axis and the boost-Hamiltonian $H.$ 

In addition there are $(d-1)$ rotations $\CJ$, which rotate the positions of the pode and antipode, and another $(d-1)$ boost operators $\CK$, that also move the pode, antipode, and horizon. I'll call these the ``hard" generators.

As an illustration here is the complete algebra for $d=3$,
\bea 
[R,H] \eq 0 \cr \cr
[R ,\CJ ] \eq  i \epsilon_{ij} \CJ_j \cr \cr
[R, \CK, ] \eq   i  \epsilon_{ij} \CK_j   \cr \cr
[\CJ_i , \CJ_j ] \eq   i \epsilon_{ij} R  \cr \cr
[\CK_i , \CK_j ] \eq   -i \epsilon_{ij}  R    \cr \cr
[ \CJ_i, \CK_j ] \eq    i \epsilon_{ij}  H  \cr \cr
[H , \CK_i ] \eq  -i \epsilon_{ij}  \CJ_j  \cr \cr
[H , \CJ_i ] \eq    i \epsilon_{ij} \CK_j
\label{Od1}
\eea

\subsection{Four Step Protocol}\label{4step}

What follows is a schematic  protocol to test the oracle's claim. If it succeeds then the oracle's  purported dual  is legitimate. If it fails then we know he is a phony.

Actually there is a middle ground. Suppose the protocol succeeds to some very high level of accuracy, but beyond that it fails. Then the right thing to do is to see if the small violations have a plausible gravitational explanation. As we will see this is not just an academic possibility.

The protocol is probably consistent at the semiclassical level---in other words to all orders in perturbation theory. Beyond that, at the level of exponentially small non-perturbative effects, the protocol must fail, but there is a plausible gravitational explanation for the failure involving higher genus\footnote{I will use the terms \it higher genus \rm and \it wormhole \rm rather loosely to mean any connected geometry with topology different from the original semiclassical geometry. For Euclidean dS the semiclassical geometry is a 4-sphere. Any other connected topology is by definition a higher genus wormhole.} saddle points in the gravitational path integral.

To formulate the protocol we first break the de Sitter symmetry by choosing a pair of static patches, the pode and the antipode. This is not a real symmetry breaking; in the semiclassical theory  it is gauge fixing\footnote{    See however  section \ref{Sec: implications}.      }. If the theory is gauge invariant then the results of any physical calculation should not depend on the gauge, which in this case means the particular pair of asymptotic points used to define the static patch.

\begin{enumerate}
\item  Step one begins with a candidate   for the dual of the  static patch.  It  consists of a conventional quantum system: a Hilbert space $\CH_{\p}$ of states; a Hamiltonian $H_{\p}$; and a set of Hermitian observables including the easy generators of the subgroup  that hold fixed the pode. These generators include the Hamiltonian $H_{\p}$ and the rotation generators $R_\p$
that rotate the static patch around the pode. The easy generators  are collectively denoted $G_\p.$

The system is assumed to be in thermal equilibrium at some definite  temperature\footnote{With the normalization of the Hamiltonian defined by \ref{Od1} the temperature of the static patch has the value $T=1/2\pi.$} $T$ and entropy $S.$
\item Step two introduces another copy of  the system labeled  $\a$ (for antipode). The doubled system has Hilbert space 
\be
\CH = \CH_{\p}\otimes \CH_{\a}.
\ee
The total Hamiltonian is,
\be 
H = H_{\p} -H_{\a}.
\ee
The Hamiltonian $H_{\a}$ is identical to $H_{\b}$.  The full Hamiltonian generates a boost that translates one side of the Penrose diagram (the pode side) upward, and the other side downward.

The full rotation generators acting on $\CH$ are
\be 
R= R_{\a} +R_{\b}.
\ee
More generally,
\be 
G=G_{\a} \pm G_{\b},
\ee
the minus sign being chosen for the Hamiltonian.

Because \dS is spatially closed without a boundary we must impose  gauge constraints,
\be 
G |\Psi\ra =0
\label{easygauge}
\ee
on the physical states.

So far none of this is unusual and can easily be satisfied in many ways.  That's why I call $G$ the ``easy" generators. Another way to characterize them is that they are the generators that commute with the Hamiltonian. And finally the easy generators  do not couple the pode and antipode degrees of freedom.

\item The third step involves the construction of the remaining ``hard" generators of $O(d,1)$, those that displace the pode and antipode. 
Let us call them $\CG.$ They consist of the remaining rotation generators $\CJ$ and an equal number of boosts $\CK$. The hard generators couple the pode and antipode degrees of freedom non-trivially.
The easy and hard generators together  form the $O(d,1)$ algebra. 

This third step may not be possible. There may be no choice of $\CG$ that satisfies the commutation relations \ref{algebra}. This in itself  may not be fatal if the algebra can be realized to a sufficiently high degree of accuracy. For example the violations may be  exponentially small $\sim \exp{(-S_0)}$.  The  symmetry may be satisfied to leading order in an expansion in $e^{-S_0}$, with the violations occurring only in higher orders.  We will see in section \ref{Sec: Od1anomaly} that this  is exactly what the GKS anomaly tells us must happen. But for now let us assume the $\CG$ can be constructed.

\item The final step, assuming the others have been successful,  is to impose the hard gauge constraints,
\be 
\CG |\Psi \ra = 0.
\label{hardgauge}
\ee
Note that the hard gauge constraints \ref{hardgauge} automatically imply the easy gauge constraints \ref{easygauge} but not the other way. As in step 3, if the entropy is finite this may only be possible to exponential precision.

\end{enumerate}

If there are no states satisfying the gauge constraints, 
\bea 
G|\Psi\ra \eq 0    \cr \cr
\CG |\Psi\ra \eq 0,
\eea 
then we stop, go back, and choose another candidate until we find one that has at least one gauge-invariant state.

The  gauge invariant state that is crucial is the de Sitter vacuum
 which looks thermal to observers in the pode and antipode patches. This state is the thermofield-double,
\be 
|TFD\ra = \sum_i e^{-\frac{\beta E_i}{2}} |E_i\ra_{\a} |{\bar E}_i\ra_{\b}.
\label{TFD}
\ee
which must satisfy the gauge constraints, at least to leading order in $\exp{(-S_0)}$. I don't see any reason why there should be other gauge invariant states but this seems to be controversial. In any case the entire discussion in   this paper is about the state in \ref{TFD}.

 We will return to these symmetry issues but first I want to digress and describe a toy model which can provide a source of intuition about de Sitter static patches.

\section{Toy Model} \label{Sec: toy}

\begin{figure}[H]
\begin{center}
\includegraphics[scale=.3]{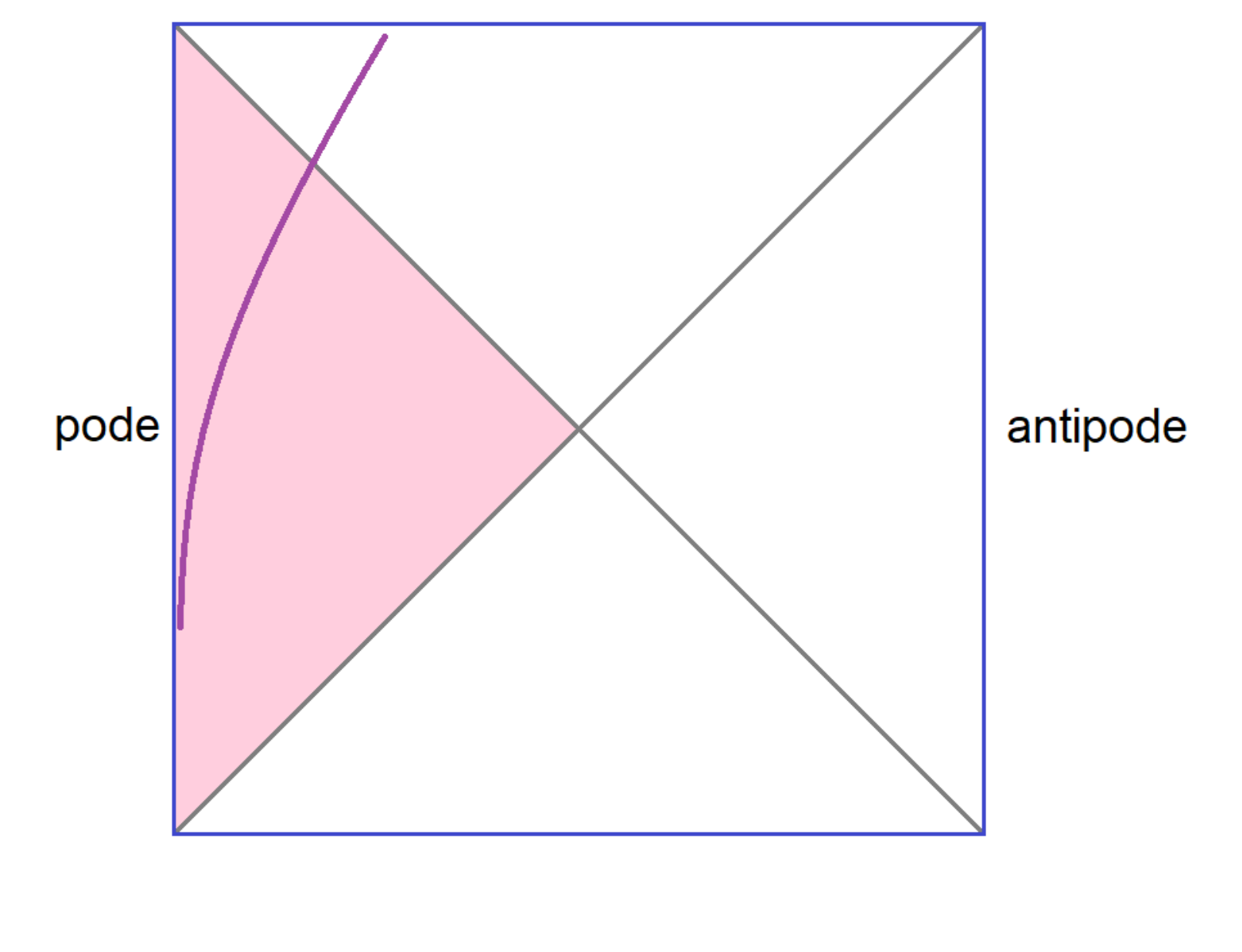}
\caption{The pode is a point of instability. A particle arbitrarily close to the pode will depart (exponentially) from the pode and eventually it will fall through the horizon. Classically it will take a logarithmically infinite coordinate time to reach the horizon.}
\label{gbye}
\end{center}
\end{figure}

 The motivation for the toy  model is the observation that the pode is a point of unstable equilibrium. Imagine a light test-particle located exactly at the pode. Consider a second test-particle a tiny distance from the pode. Assume the second particle is initially at rest relative to the pode and subsequently follows a geodesic. Geodesic deviation will cause that particle to fall away from the pode with the separation growing exponentially. This is illustrated in fig \ref{gbye}.
To model this behavior we can consider a non-relativistic particle in an inverted three-dimensional harmonic oscillator potential,
\be 
H=\sum_{i} \frac{p_i^2-x_i^2}{2}   \ \ \ \ \ \ \ \ \ (i=1,2,3)
\label{H}
\ee
The unstable equilibrium  point $x_i=0$ corresponds to the pode. 
The spectrum of $H$ is continuous and runs over all real numbers. The energy eigenfunctions at large $x$ have the form,
 \be 
 \psi \to e^{\pm i x^2/2}.
 \label{WF}
 \ee

Classically, if the particle starts near the top of the potential the subsequent motion satisfies,
\bea
r&\equiv&\sqrt{|x|^2} \to e^{t} \cr \cr
|p |&\to& e^{t},
\label{exptau}
\eea
where $p$ is the momentum of the particle. This matches the behavior of a particle in dS.

The time that it takes for the particle to get to distance $R$ from the pode is,
\be 
t_* \approx \log R.
\label{scrambt}
\ee
I've intentionally used the notation $t_*$ which is the conventional notation for the scrambling time. The reason will become clear shortly.

The inverted oscillator is characterized by an operator algebra including the Hamiltonian, and  for each direction a generator $L_{\pm}$ defined by,
\be 
L_{\pm} = \frac{x \pm p}{\sqrt{2}}.
\label{Lpm}
\ee
The algebra, which I will call the \it symmetry  \rm of the model is,
\bea 
[H, L_-] &=& iL_-   \cr \cr
[H, L_+] &=& -iL_+ \cr \cr
[L_-, L_+] \eq i.
\label{algebra}
\eea
From the first two of these relations it follows that,
\bea 
L_-(t) \eq L_- e^{-t} \cr  \cr
L_+(t) \eq L_+ e^{t} 
\label{exptau2}
\eea

The algebra is satisfied by a generalization to a system of many non-interacting particles as well as particles coupled by translationally invariant forces. 

The toy model as defined up to this point is the semiclassical limit of a more complete model which  has a stretched horizon, a finite entropy, and nonperturbative effects.

\subsection{Toy Model with Stretched Horizon}

So far, in the  toy model the particle falls in the potential for an infinite amount of time before reaching $r=\infty$. This  parallels the fact that classically, a particle takes  infinite  time to reach the de Sitter horizon. 

To make a more interesting model the radial direction can be cut off by turning the potential sharply upward at a distance $R$ from the pode. Instead of a single particle we can introduce $N$ particles which are allowed to interact, but only when they are very near the bottom of the potential. The details of the interaction are not important other than they lead to chaotic behavior and thermalization.

\begin{figure}[H]
\begin{center}
\includegraphics[scale=.4]{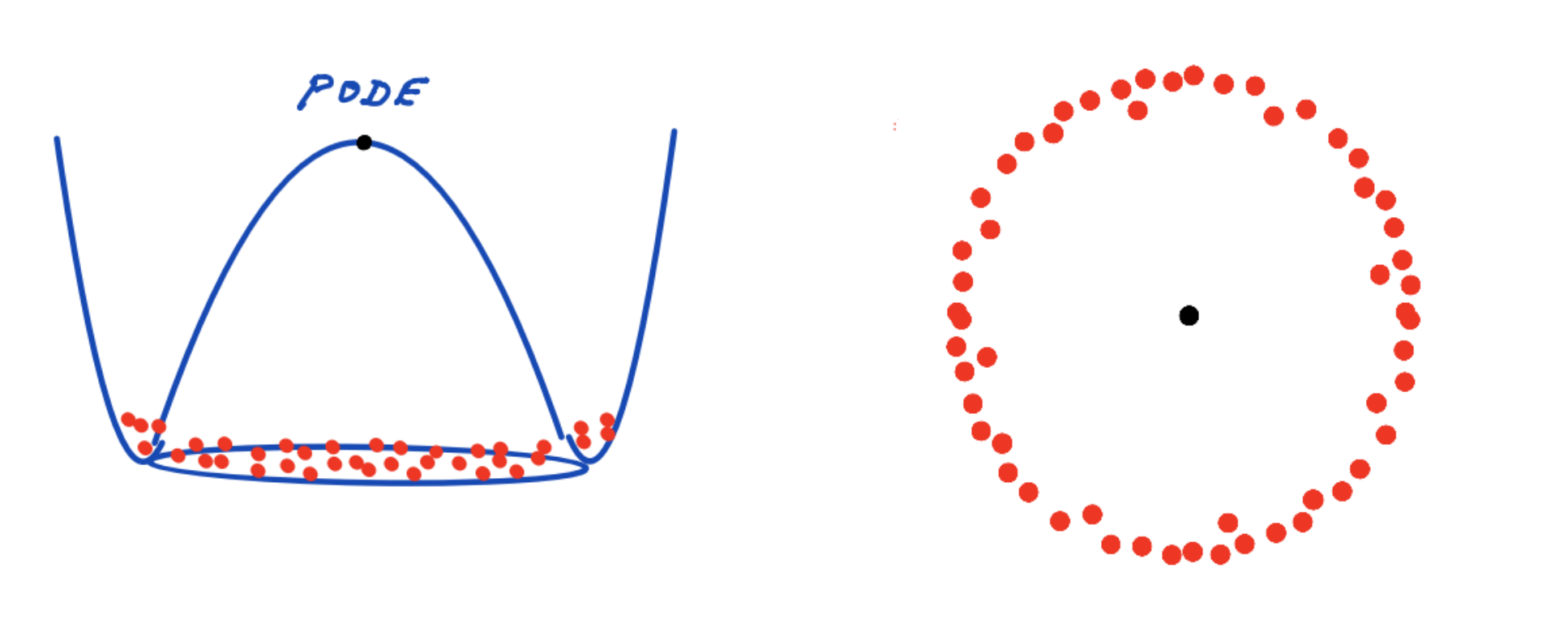}
\caption{Toy model with stretched horizon. The right panel shows the view from above.}
\label{invert}
\end{center}
\end{figure}

Figure \ref{invert} illustrates the setup. The first panel shows the potential, along with $N$ particles in thermal equilibrium at the bottom. The second panel is the view-from-above in which we see the particles occupying a  two-dimensional  shell at distance $R$ from the pode. This shell is the toy model version of the stretched horizon.

The number of particles can be chosen so that the area-density  at the stretched horizon is  of order $1/G.$ The entropy of the thermal gas is proportional to the number of particles and and by an appropriate choice of temperature and other  numerical constants,
\bea 
S_0  \eq \frac{\rm Area}{4G} \cr \cr
\eq \frac{\pi R^2}{G}
\label{S=A}
\eea
Note that the time that it takes for a particle to fall from the pode to the stretched horizon is  the scrambling time.

The semiclassical limit is  identified with the limit  $R\to \infty.$ In that limit the entropy becomes infinite and the thermal state is very boring to an observer at the pode. 
Thermal fluctuations do occur, but the
potential well is so deep that the probability for a particle in the equilibrium ``soup"  to reach the pode is zero. 

\subsection{Fluctuations}

If $R$ (and therefore the entropy) is kept finite, there is a non-zero probability to find one or more particles at the pode. The particles could form  interesting objects such as  black holes,  galaxies, or  brains. These nonperturbative Boltzmann fluctuations are extremely rare, but they are the only things that happen in thermal equilibrium \cite{Dyson:2002pf}.

Consider the probability of a Boltzmann fluctuation in which an object $\CO$ materializes at the pode. To represent this situation mathematically we introduce a projection operator $\Pi_{\CO}$ that projects onto states in which the object $\CO$ is present at the pode. The probability in question is given by,
\be 
 P_{\CO} = Tr \  \rho \Pi_{\CO}
 \label{P=Tr}
\ee
where $\rho$ is the thermal density matrix.

Non-equilibrium dynamics is encoded in the thermal state. For example, suppose we want to know the probability that if the object $\CO$ is present at $t=0,$ then at a later time $t$ it will have made a transition to $\CO'.$ This is encoded in the correlation function,
\be 
Tr \rho \ \Pi_{\CO}  \  e^{-iHt} \ \Pi_{\CO' } \ e^{iHt}
= Tr \rho \ \Pi_{\CO}(0) \ \Pi_{\CO' }(t)
\ee
This type of formula, and generalizations of it, show that the theory of fluctuations in thermal equilibrium encodes a very rich spectrum of dynamical phenomena.

The probability $ P_{\CO}$ is given by a standard expression,
\be 
 P_{\CO} = e^{-\Delta S}
 \label{Prob}
\ee
where the ``entropy-deficit"  \ $\Delta S$ is defined  by, 
\be 
\Delta S = S_0 -S_{\CO}.
\label{Delta}
\ee
This formula requires some explanation. The symbol $S_0$ stands for the de Sitter entropy $\pi R^2/G.$  $S_{\CO}$  however, does not stand for the entropy of the object $\CO.$    It is the conditional  entropy of the whole system, given that the object $\CO$ is present at the pode. 
A simple  way to think about $S_{\CO}$    is that it  represents  the entropy of the remaining horizon degrees of freedom, given that $\CO$ is present, plus the entropy of $\CO.$ We will return to this in section 
\ref{Sec: UseGR }.

\subsection{The GKS Anomaly} \label{Sec: Anomaly}

Can the algebra \ref{algebra}  be satisfied in the cutoff model? With the identification \ref{Lpm} and the modification of the Hamiltonian required to construct the cutoff model, the algebra will not hold, but one may ask if there can be new operators $L_{\pm}$, which along with the new Hamiltonian, satisfy it? The answer is no \cite{Goheer:2002vf}. To prove it\footnote{A more rigorous version of the proof was given in \cite{Goheer:2002vf}. } we consider the first of equations \ref{exptau2} (a consequence of the algebra)  and take its matrix element between normalizable states,
\be 
\la \psi| L_-(t) |\psi \ra \to e^{-t} \to 0.
\label{nogo}
\ee
The argument for the GKS anomaly contains two parts:

\begin{enumerate}

\item  Finiteness of entropy implies that the energy  spectrum is discrete. More exactly it says that the number of states below any given energy is finite. This is much weaker than saying the Hilbert space is finite dimensional, which we  do not assume.

\item  Functions defined as sums of the form,
\be 
F(t) = \sum_i a_i e^{iE_i t}
\label{sum}
\ee
cannot go to zero as $t \to \infty.$ They will have fluctuations and even recurrences on very long time scales. One can easily prove that the late-time variance of $F(t)$ satisfies,
\be 
\lim_{T\to \infty} \frac{1}{T} \int_0^T |F^2| dt  = \sum_i |a_i|^2 >0
\label{lim}
\ee
In other words, over long periods of time $F$ will fluctuate with a variance equal to 
$ \sum_i |a_i|^2 .$ 
\end{enumerate}
It follows  \cite{Goheer:2002vf} that \ref{nogo}  (and the algebra \ref{algebra} which led to it)  cannot be satisfied if the entropy is finite.
There is a deep relationship between fluctuations and the non-perturbative breaking of semiclassical symmetries.

We can be more quantitative. In the semiclassical limit the energy spectrum is continuous and \ref{sum} is replaced by,
\be 
F(t) = \int A(E) e^{iEt}.
\ee
If $A(E)$ is square integrable then $F(t) \to 0$ as $t\to \infty.$
In approximating the sum by an integral we make the following correspondence:
\be 
a_i =A(E) \delta E
\ee
where $\delta E$ is the spacing between neighboring energy levels.
Now consider the sum in \ref{lim} and rewrite it in terms of $A$.
\bea
\sum |a_i|^2 \eq \sum |A(E)|^2 (\delta E)^2 \cr  \cr
&\to&  \int |A(E)|^2  \delta E dE  
\label{spread}
\eea

The energy level spacings $\delta E$ are of order $e^{-S_0}.$ Thus it follows that 
$|\sum a_i|^2 \sim \delta E \sim e^{-S_0},$ and from \ref{lim},
\be
\rm Var \it(L_-) \sim e^{-S_0}.  
\label{Variance}
\ee
Given that the symmetry algebra requires the asymptotic variance in $L_-$ to be zero, the actual variance in \ref{Variance} is a measure of how badly the symmetry is broken by the 
anomaly.

More generally it seems reasonable to suppose that the discreteness of the energy spectrum produces effects that scale like a power of $e^{-S_0}.$ 

The bottom line is that fluctuations of order $e^{-S_0}$   create an obstruction to realizing the symmetry algebra
 \ref{algebra} as long as the entropy is finite.

\subsection{Caveats}
The toy model has elements in common with de Sitter space, but like all analogies it has limitations. Two come to mind:
First, because it is based on non-relativistic particles it cannot capture the physics of massless photons in de Sitter space. For example the probability to find a single thermal photon
of wavelength $\sim R/2$ within a distance $\sim R/2$ of the pode is order $1.$ This is a perturbative phenomenon which would requires massless relativistic degrees of freedom instead of massive non-relativistic particles.

Another unphysical feature of the toy model is that the size of the horizon is fixed, while the size of the de Sitter horizon is dynamical and adjusts to the amount of entropy that it contains (see section \ref{Sec: UseGR }).

\section{The GKS Anomaly in JT/SYK}
\begin{figure}[H]
\begin{center}
\includegraphics[scale=.4]{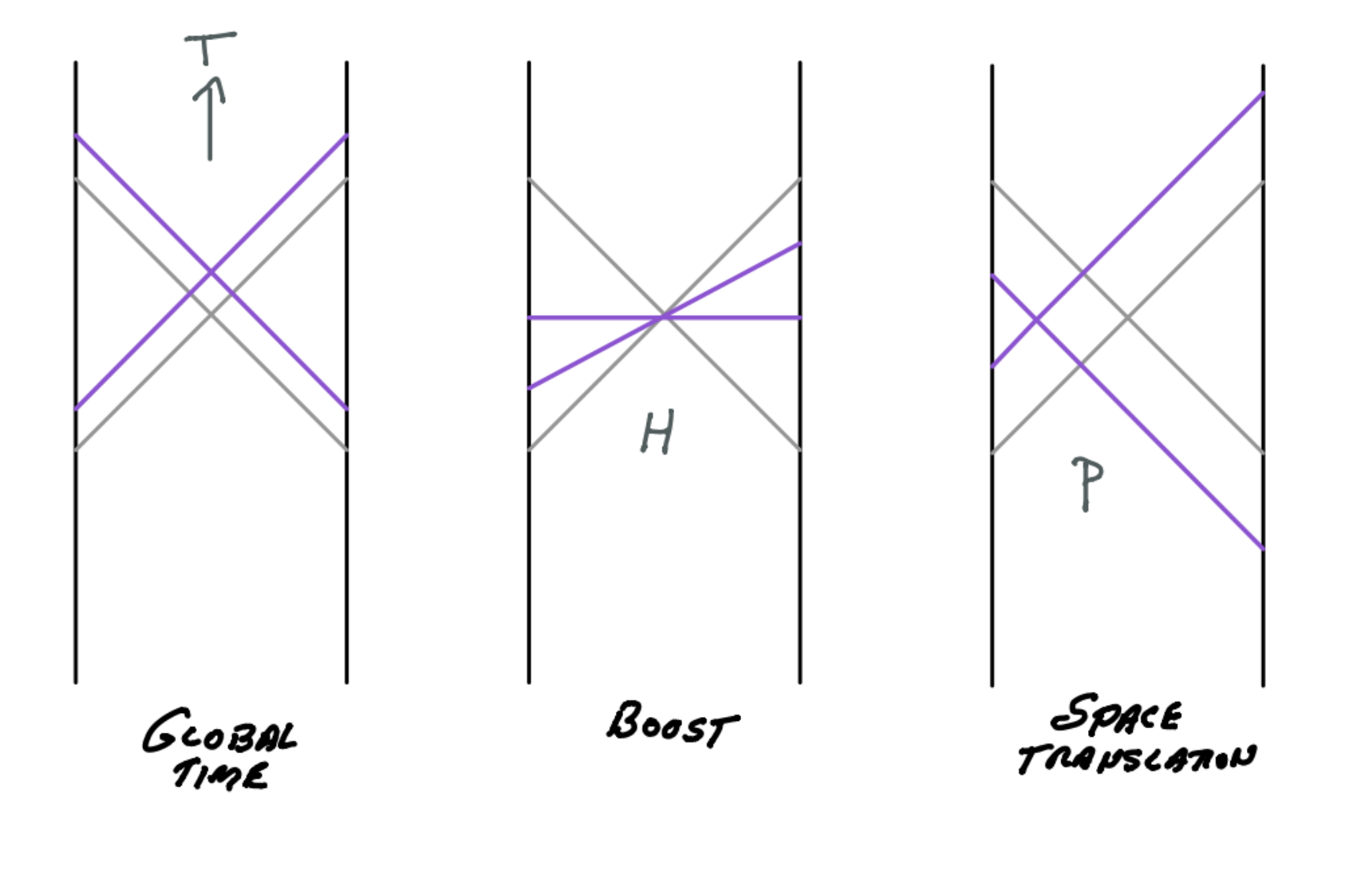}
\caption{The three generators of $SL(2R)$:  $\CT$ generates global time translations and shifts the horizon vertically; $H$ is the boost Hamiltonian that generates Rindler-like boosts; and $\CP $ generates space-like shifts of the horizon. Only $H$ does not move the horizon.}
\label{ads2}
\end{center}
\end{figure}

A similar anomaly to that seen in section
  \ref{Sec: Anomaly}
 also affects two-dimensional models such as JT gravity and  its                                                                                                 quantum   SYK completion.  
Figure \ref{ads2} shows  the $AdS(2)$ solution of JT gravity  with its Rindler-like horizons.
\begin{figure}[H]
\begin{center}
\includegraphics[scale=.4]{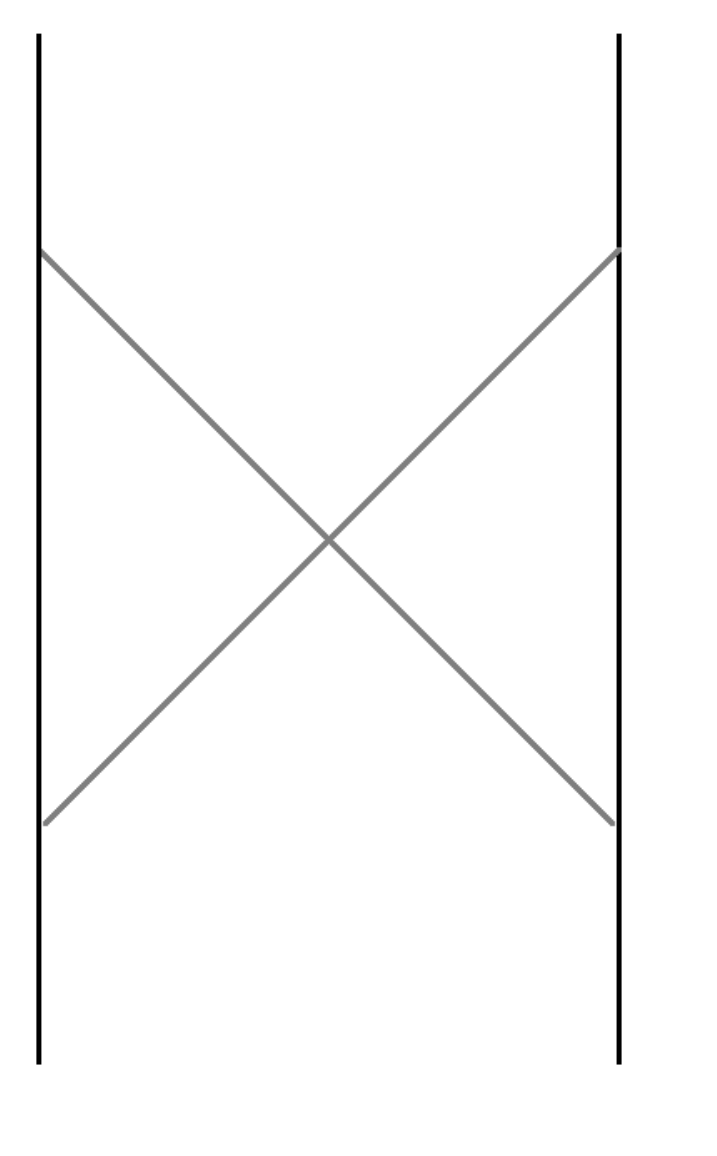}
\caption{$AdS(2)$ with horizons.}
\label{ads}
\end{center}
\end{figure}
The classical theory\footnote{JT gravity coupled to matter with the matter being uncoupled to the dilaton. I am grateful to Douglas Stanford and Henry Lin for explaining this to me.} has an exact $SL(2R)$ symmetry which  persists to all orders in perturbation theory; in other words it is a feature of the semiclassical theory.  However the symmetry is broken by non-perturbative quantum effects  \cite{Maldacena:2016upp}.
The $SL(2R)$ group has three generators, $\CT,H$ and $\CP,$ whose action is
 illustrated diagramatically \cite{Lin:2019qwu} 
 in the three panels of fig \ref{ads2}.
\bn
The generator $\CT$ generates global time shifts and moves the horizons vertically from their original location (grey lines) to a new location (purple lines.) The generator $H$ is the boost Hamiltonian that holds the horizons fixed but acts on equal-time-slices to boost them. Finally $\CP$  generates a spacelike displacement of the horizon as shown. $\CT$ and $\CP$ are analogous to the hard generators of $O(d,1)$ and $H$ is easy. 

These generators satisfy the $SL(2R)$ algebra,
\bea
[H, \CT] \eq i\CP \cr \cr
[H,\CP] \eq i\CT \cr  \cr
[\CT,\CP] &=&iH
\label{SL2R}
\eea
which as I said is exact semiclassically, but cannot hold non-perturbatively. To see this we define the light-like generators,
\bea
L_- \eq \frac{\CP-\CT}{\sqrt{2}} \cr \cr
L_+ \eq \frac{\CP+\CT}{\sqrt{2}}.
\label{LL gen}
\eea
From \ref{SL2R},
\be 
i[H, L_{\pm}] = \pm L_{\pm}
\label{[HL]}
\ee
implying,
\be 
 L_{\pm}(t) =  L_{\pm}e^{\pm t}.
 \label{L=epmt}
\ee

 In the classical JT system the entropy is infinite and equation \ref{L=epmt} presents no problem, but in a quantum completion such as SYK the entropy is finite, of order the number of fermion species $N.$  The rest of the argument is identical to the one in section \ref{Sec: Anomaly} and
 implies that the $SL(2R)$ symmetry cannot be exact, except in the limit $N\to \infty.$  

All of this is well known from other points of view \cite{Maldacena:2016upp}  \cite{Lin:2019qwu}, and it is believed that the breaking of the symmetry can be understood in terms of higher genus corrections to the JT path integral.

\section{de Sitter} \label{dS}
Return now to de Sitter space and  static patch holography. In so far as the static patch is in thermal equilibrium with  finite entropy it will undergo Boltzmann fluctuations. As we saw in section   \ref{Sec: Anomaly} these fluctuations are the source  the symmetry breaking in the toy model. In this section I'll do two things: first  I'll discuss the anomaly in the $O(d,1) $ de Sitter symmetry;  and  then I'll explain how general relativity can be used to give a quantitative account of Boltzmann fluctuations

\subsection{The O(d,1) Anomaly} \label{Sec: Od1anomaly}
Now let's return to the symmetry algebra of de Sitter space.
 From the last two of  equations \ref{Od1} we may construct light-like generators,
\be 
 L_\pm =\frac{\CJ_2 \pm \CK_1}{\sqrt{2}}
 \ee
 satisfying,
 \be 
i[H, L_{\pm}] = \pm L_{\pm}.
\label{[HLds]}
\ee
Note that these equations are identical to the first two equations in \ref{algebra} and that they again imply,
\be 
L_-(t) = L_-(0) e^{-t}.
\ee
Following the same logic as in section \ref{Sec: Anomaly} we may evaluate this equation between normalizable states to find,
\be 
\langle \psi | L_-(t) |\psi\ra \to 0.
\label{uh oh}
\ee
For the same reasons   as in  \ref{Sec: Anomaly} and in \cite{Goheer:2002vf} (having to do with persistent fluctuations) there is a GKS anomaly making  it  impossible to satisfy \ref{uh oh}, and therefore   the algebra. The  arguments of section \ref{Sec: Anomaly} concerning fluctuations    give the same estimate for the magnitude of  the effects of the anomaly,
\be 
\rm Var \it(L_-) \sim e^{-S_0}. 
 \label{variance}
\ee

\subsection{Using GR to Calculate Fluctuation Probabilities} \label{Sec: UseGR }

In theories with a gravitational dual, general relativity provides a precise  way of calculating the probability for certain fluctuations.
Consider the rate for  fluctuations which nucleate  massive objects such as  black holes near the pode. We may make use of \ref{Prob} and \ref{Delta} to compute the rate. 
To calculate $S_0$ we can use the metric \ref{SPmetric} to get the radius of the horizon (we get $r_H = R$), then calculate the area of the horizon, and finally the entropy. The result is,
\be 
S_0=\frac{\pi R^2}{G}.
\label{entr}
\ee

Next consider the metric of de Sitter space with an object $\CO$ of mass $M$ at the pode. The effect of the mass is to pull in the cosmic horizon, shrinking its area and therefore its entropy  \cite{Banks:2006rx}\cite{Susskind:2011ap}\cite{Banks:2016taq}.  Out beyond the radius of the object the metric takes the form \ref{SPmetric} except that the emblackening factor $f(r)$ is replaced by
 \be
f_M(r) = \lf 1- \frac{r^2}{R^2 } -\frac{2MG}{r} \rg.
\label{BHmetric}
\ee

The horizon location is defined by $f_M(r) =0.$ Multiplying by $r$ this becomes,
\be  
r-\frac{r^3}{R^2} -2MG =0
\label{horiz-eq}
\ee

Equation \ref{horiz-eq} has three solutions, two with positive $r$ and one with negative $r.$
The negative solution is unphysical. The larger of the two positive solutions determines the  location of the cosmic horizon and the smaller determines the  horizon of a black hole of mass $M$. To lowest order in $M$ the 
cosmic horizon is shifted to a new value of $r$ given by,
\be 
r = R-MG
\label{shift}
\ee
The entropy is given by,
\be 
S_{\CO} = \frac{\pi (R-MG)^2}{G}
\label{newA}
\ee
and (to leading order in $M$) the entropy-deficit by   \cite{Banks:2006rx},
\be 
\Delta S = 2\pi MR.
\label{D=2piMR}
\ee

Now recall that the inverse temperature of \dS is $\beta =2\pi R.$ Using \ref{Prob} we find,
\be 
P_{\CO} = e^{-\beta M}.
\label{Boltz}
\ee
Equation \ref{Boltz} is the Boltzman weight of the for a configuration of energy $M.$ The answer itself is not surprising but what is interesting is that the connection between entropy and area has been used in a new way---not for  equilibrium probabilities but for fluctuations away from average behavior.

More generally we can go past linear order in $M$. Let us denote the two solutions of \ref{horiz-eq} by $r_-$ and $r_+$ and define the independent  parameter,
\be 
x=(r_+ - r_-).
\ee
One can express $x$ in terms of the mass of the black hole by eliminating $r_+$ and $r_-$ from the equations,
\bea
R^2 \eq r_+^2 +r_-^2 +r_+r_-  \cr 
  2MG R^2   \eq  r_+r_-(r_+ + r_-)   \cr 
  x \eq r_+-r_-
\eea

The value of $x$ runs from $(x=-R)$ to $(x=+R).$  Changing the sign of $x$ interchanges the cosmic and black hole horizons. It is convenient to think of positive and negative values of $x$ as different configurations; for example at both $x=-R$ and $x=+R$
 there is a vanishingly small  horizon and a maximally large  horizon,  but we regard the two states as different. 
 
The entropy and entropy-deficit are given by
\bea 
S_\CO \eq \frac{\pi}{G} \lf r_+^2 +   r_-^2           \rg  \cr \cr
\Delta S \eq  \frac{\pi}{G} \lf R^2 - r_+^2 -   r_-^2           \rg
\label{SandDS}
\eea

Let us consider the ``Nariai point"  $(x=0)$ at which $(r_+ = r_-).$  One easily finds that at the Nariai  point,
\bea 
r+^2 \eq r_-^2   \cr \cr
&\equiv&   r_N^2  \cr \cr
\eq \frac{R^2}{3}
\label{N-pt}
\eea
 From \ref{SandDS} and \ref{N-pt} we find,
 \bea
 S_N \eq \frac{2S_0}{3} \cr \cr
 \Delta S_N \eq \frac{S_0}{3}
 \label{DSN}
 \eea

\begin{figure}[H]
\begin{center}
\includegraphics[scale=.4]{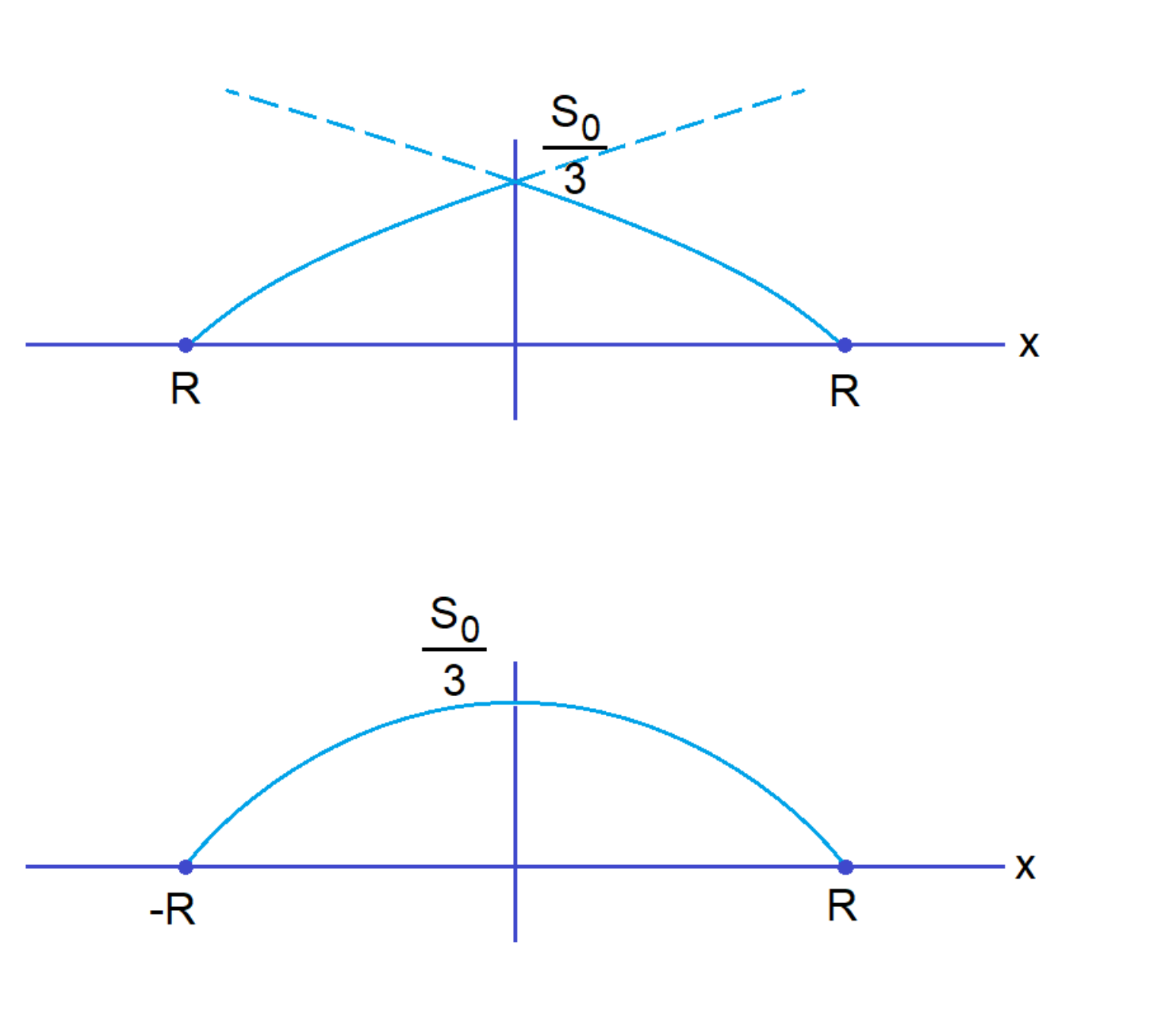}
\caption{Entropy deficit and a function of $x$. The upper panel incorrectly  shows a cusp at the Nariai point. The lower panel shows the correct smooth behavior \ref{smooth}. }
\label{sx}
\end{center}
\end{figure}

It is not obvious that $\Delta S$ is smooth at $x=0$. One might expect that $\Delta S(x)$ has a cusp as in the top panel of fig \ref{sx}. However explicit calculation shows that the dependence is completely smooth and surprisingly simple,
\be 
\Delta S= \frac{S_0}{3}  \lf  1 - \frac{x^2}{R^2}    \rg.
\label{smooth}
\ee
This is  illustrated in the bottom panel of fig \ref{sx}.

The total probability for a black hole fluctuation is given by an integral\footnote{The factor $x^3$ in the integrand of \ref{integral} is due to three negative modes associated with the instability of an object located at the pode. See
figure \ref{gbye} and equation \ref{H}.}
over $x$,
\bea 
\rm Prob \it \eq    \frac{1}{R^4}  \int_0^R  e^{-\Delta S}   x^3dx \cr \cr
\eq e^{\frac{-S_0}{3}}\frac{1}{R^4} \int_0^R  e^{\frac{S_0x^2}{3R^2}} x^3 dx.
\label{integral}
\eea
One finds,
\be 
\rm Prob \it \sim \frac{3G}{\pi R^2 } \lf   1 + \frac{3}{S_0}    e^{-S_0/3}    \rg
\label{prob}
\ee

The first term in \ref{prob} comes from the endpoint of the integration at $x^2 =R^2$ and is perturbative in $G.$ It is associated with the lightest black holes and numerically dominates the integral. The second term comes from the saddle point at $x=0.$ It
can be re-written as 
$$e^{-\frac{\pi R^2}{3}\frac{1}{G}}.$$ It is obviously non-perturbative in $G.$ We will discuss its meaning in the next section.

\section{The Nariai Geometry} \label{Sec: high genus}

Quantum mechanically  the non-perturbative fluctuations  we are discussing originate from  the discreteness of the energy spectrum. On the gravitational side those same effects are  encoded in higher genus\footnote{See footnote 3 in section \ref{4step}.}  contributions to the gravitational (Euclidean) path integral \cite{Cotler:2016fpe}\cite{Saad:2018bqo}\cite{Saad:2020bvb}\cite{Marolf:2020xie}\cite{Almheiri:2019hni}. 
In the case of anti de Sitter space the geometries that contribute are constrained to have asymptotic  AdS-like boundary conditions, but subject to that constraint they can have any topology. 
In the case of  Euclidean de Sitter space there are no boundaries; the  path integral therefore includes all closed topologies. Among these are the Nariai geometries.

\begin{figure}[H]
\begin{center}
\includegraphics[scale=.5]{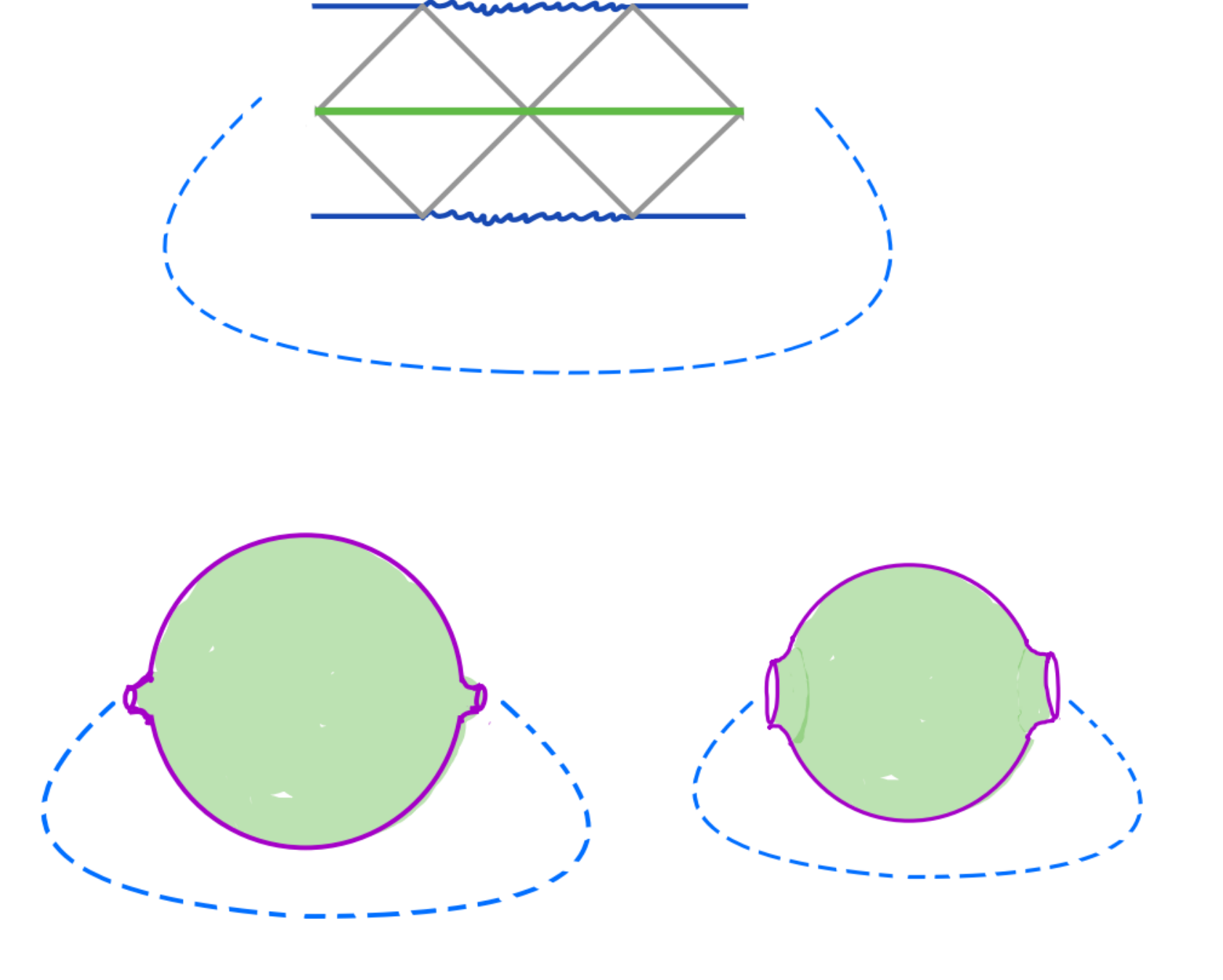}
\caption{de Sitter-Schwarzschild black hole. The upper panel shows the Penrose diagram and the two lower panels show the spatial geometry of time-symmetric slices.  In each case the dashed blue curves indicate geometric  identifications.}
\label{nucleate}
\end{center}
\end{figure}

To illustrate the connection between fluctuations and higher-genus contributions consider 
the fluctuations discussed in section \ref{Sec: UseGR } in which a black hole is spontaneously created near the pode.
The first panel of figure \ref{nucleate} shows the Penrose diagram for a \S- de Sitter black hole.
A slice through the time-symmetric space-like $t=0$ surface is depicted in green. In the lower panel the geometry of such a slice is shown for two different masses of the black hole. The figure on the right depicts a relatively larger mass than on the left.
\bn

One may continue the $t=0$ geometry in either Minkowski or Euclidean signature. The Minkowski continuation just gives back the geometry in the top panel of figure \ref{nucleate}. The Euclidean continuation gives a compact  geometry with the topology
$S_2\times S_2$ which I will call an S-geometry (Euclidean \S \ de Sitter). 
There is a one parameter family of S-geometries parameterized by the mass $M$ of the black hole.  

The metric of the S-geometry is,
\be  
ds^2 =\lf 1-\frac{r^2}{R^2}  -\frac{2MG}{r} \rg d\t^2 + \lf 1-\frac{r^2}{R^2}  -\frac{2MG}{r} \rg^{-1}dr^2  +r^2 d\Omega^2
\ee

The one thing left to specify is the range of the periodic Euclidean time $\tau.$ Normally the  periodic constraint requires $0<\tau\leq \beta$ where $\beta$ signifies inverse temperature. However in the present case there is no well defined temperature because the black hole and cosmic horizon have very different temperatures. For a small black hole $\beta$ is $4\pi MG$ while the de Sitter value of $\beta$ is $2\pi R.$ The black hole is far out of equilibrium with the de Sitter space.

What this means geometrically is that it is not possible to avoid a conical singularity at either the black hole or the cosmic horizon.  For this reason the S spaces are not genuine saddle points of the Euclidean gravitational path integral---with one exception. The exception is the Nariai space, i.e., S-space at the symmetric point $x=0.$ The geometry of Nariai space is $S_2 \times S_2$ and its Minkowski continuation is $S_2 \times dS_2,$ i.e., a spatial sphere times two-dimensional de Sitter space.

\begin{figure}[H]
\begin{center}
\includegraphics[scale=.3]{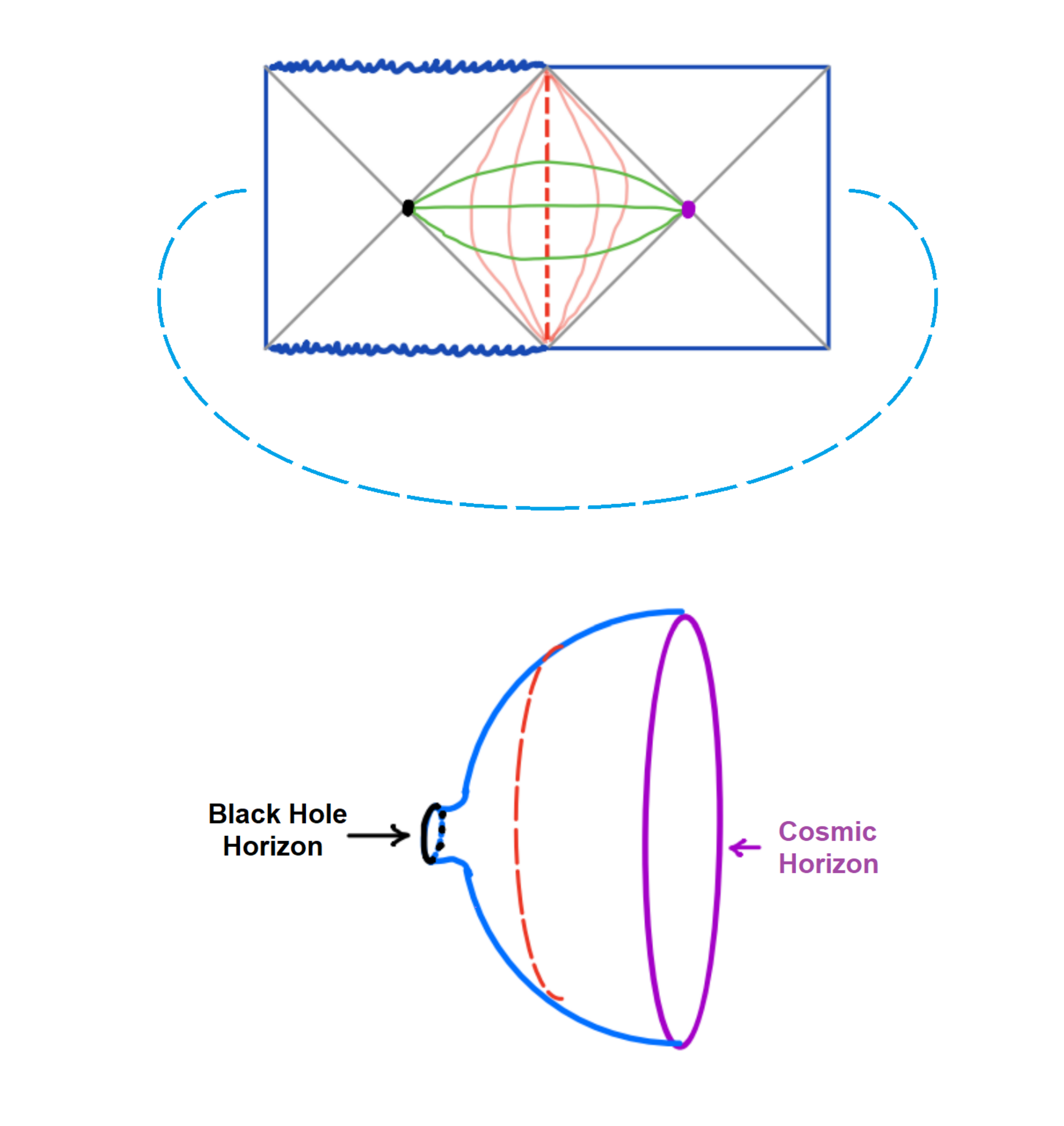}
\caption{The top panel shows the penrose diagram for a Euclidean Schwarzschild black hole together with  a static patch which encloses the black hole symmetrically. The black dot on the left is the black hole horizon.  The purple dot on the right is the cosmic horizon.   The lower panel shows a spatial slice of the region between the two horizons. The dashed red line indicates the static  position of a $2$-sphere  somewhere between the two horizons. }
\label{Sspace}
\end{center}
\end{figure}

Nariai space is a genuine Euclidean saddle point whose Minkowski continuation is often called the Nariai black hole. To understand it better let's go back to S-space with a small black hole.  A static patch is also shown in the upper panel of fig \ref{Sspace}. Because the black hole is at the center of the static patch we cannot think of the pode as a point. Instead I've indicated a shell by a dashed red line. The shell is a 2-sphere that surrounds the black hole between the black hole horizon (black dot) and the cosmic horizon (purple dot).

A spatial $t=0$ slice of the geometry between the two horizons is shown in the lower panel.

Now let us consider deforming the geometry by increasing the black hole mass and at the same time decreasing the cosmic horizon area (following the curve in the lower panel of fig \ref{sx}) until we reach the Nariai point. At that point the geometry---not just the topology--- is $S_2 \times S_2$ but from the viewpoint of the static patch observer, it looks like a spatial interval times a $2$-sphere. The observer is sandwiched between two equal horizons.

 Thought of as a real process, this history would violate the second law of thermodynamics, but as long as $S_0$ is finite, it can happen as a rare Boltzmann fluctuation. What it requires is for energy to be transferred from the cosmic horizon to the black hole---a kind of anti-evaporation. The most likely trajectory for the system to follow is the time reverse of the evaporation process in which one of the two Nariai horizons spontaneously emits a bit of radiation which is then absorbed by the other. When this happens the emitter loses energy and becomes hotter. The result is that it emits more energy until the smaller horizon becomes a small black hole (or even no black hole) and the larger horizon reaches  entropy $S_0.$ The time reverse of this process is the Boltzmann fluctuation that leads to the Nariai state from the small black hole state.
 
\begin{figure}[H]
\begin{center}
\includegraphics[scale=.5 ]{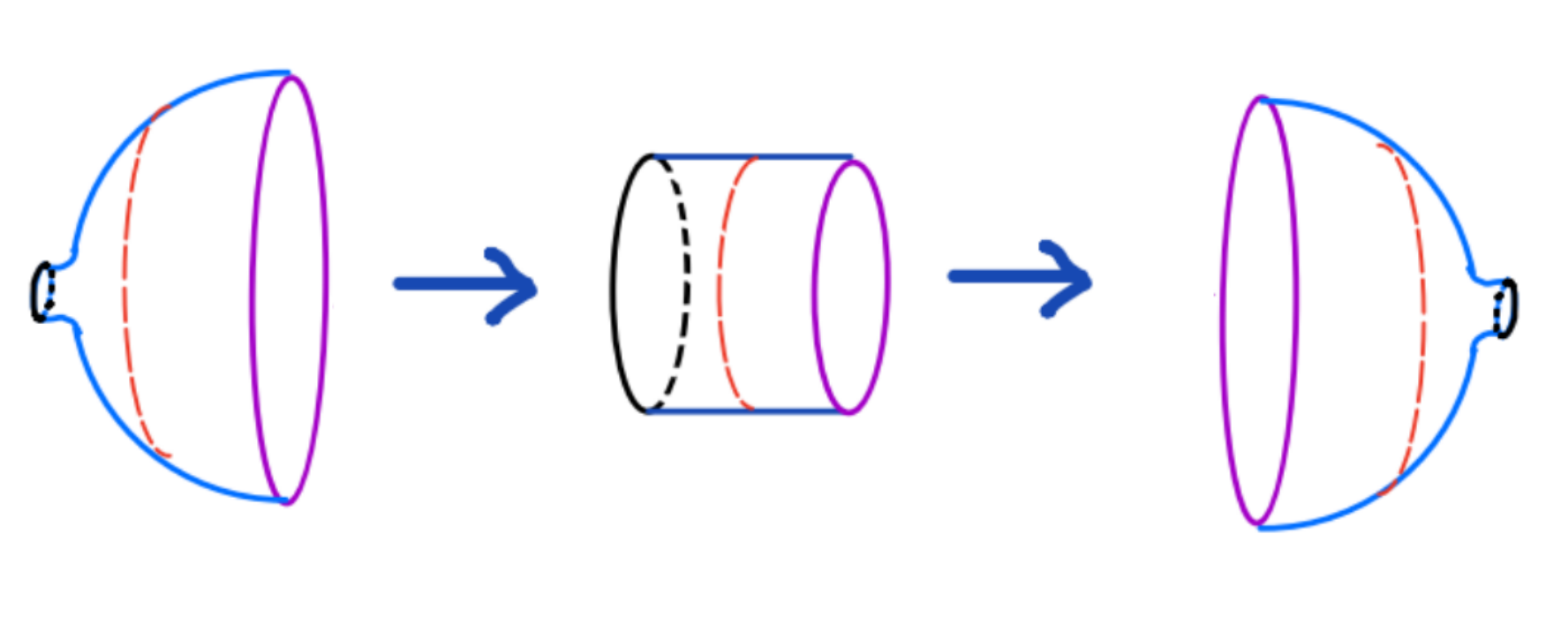}
\caption{The ``inside-out transition" in which a small black hole horizon and the cosmic horizon become interchanged by passing through the intermediate Nariai black hole.}
\label{updown}
\end{center}
\end{figure}

 Once the Nariai state is achieved it is unstable. One possibility is that the system can return to the original state with the original black hole shrinking and the original cosmic horizon returning to its full dS size. But the opposite can also happen: the system overshoots a bit and the original black hole keeps absorbing energy while the original cosmic horizon shrinks down to a small black hole. We can think of this as a transition from
 from $x=-R$ to $x=+R.$
 An observer between the two horizons sees a surprising history in which the geometry of the static patch turns itself ``inside-out"---the outer cosmic horizon and the inner black hole horizon exchanging roles. This remarkably strange event is illustrated in fig \ref{updown}.

 \subsection{Nariai and Hawking-Moss}
 
 One might be tempted to think of this inside-out transition as a quantum tunneling event, but unlike a typical quantum  tunneling the process stretches out over a long time. The energy transfer is simply Hawking evaporation or its time-reverse, and takes a time of order the Page time\footnote{One should distinguish two time scales. The first is how long the inside-out transition takes. That is a time of order the Page time. The second is the typical time between such events. This latter time scale is exponential $\sim e^{S_0/3}$. }. The transition is a thermal process  mediated by  a Hawking-Moss instanton\cite{Hawking:1981fz}, not a quantum tunneling event. It takes a very long time during which the system sits at or near the top of the potential, i.e,  at $x=0$. The Hawking-Moss instanton calculates the probability that the system in question is at the top of a broad potential barrier \cite{Weinberg:2006pc}.
 
 In this case the Hawking-Moss instanton is the Nariai geometry, $S_2 \times S_2$ and the probability to find the system in the Nariai state is $$e^{-{ \Delta  I_N}}$$ where $I$ is the action-deficit  of Euclidean Nariai space. Not surprisingly that action deficit is the same as the entropy-deficit of the Nariai black hole,
 \be 
 \Delta  I_N = \frac{S_0}{3}.
 \ee
 
 Thus we see an example of the relation between fluctuations (The appearance of a Nariai black hole) and a higher genus wormhole geometry (the $S_2\times S_2$ Nariai geometry.) 

\subsection{Connection with Anomaly}
What does all of this have to do with the $O(d,1) $ symmetry (or lack of it) of de Sitter space? I think the answer is fairly simple. In the Euclidean continuation the symmetry group is $O(5).$ There is a natural action of the $O(5) $ group on the semiclassical Euclidean dS geometry, namely $S_4$. But the full path integral receives contributions from other topologies, in particular the Nariai geometry $S_2 \times S_2.$ Trying to define the action of  $O(5)$ on $S_2 \times S_2$  is like trying to define the action of $O(3)$ on a torus. It's not that the torus breaks the symmetry like a egg would; the symmetry operations just don't exist on the torus. Likewise the generators of  $O(5)$ don't exist on $S_2 \times S_2.$

One might try to get around this by defining the action of the group to be trivial on all higher topologies; in other words define $S_2 \times S_2$ to be invariant under  $O(5).$ I think the reason that this doesn't work is that in general, states of different topology are not orthogonal: the overlaps are order $\exp(-S)$  \cite{Marolf:2020xie}. For this reason the action of the group on $S_2 \times S_2$ cannot be arbitrarily chosen independently of the action on $S_4.$

Thus we are left with the conclusion that higher topologies not only break the symmetry of dS: they don't even allow it to be defined. This is consistent with the GKS anomaly which also implies that the generators cannot consistently be constructed to order $\exp{(-S_0)}$.

\section{Implications of the Anomaly} \label{Sec: implications}
\begin{figure}[H]
\begin{center}
\includegraphics[scale=.5]{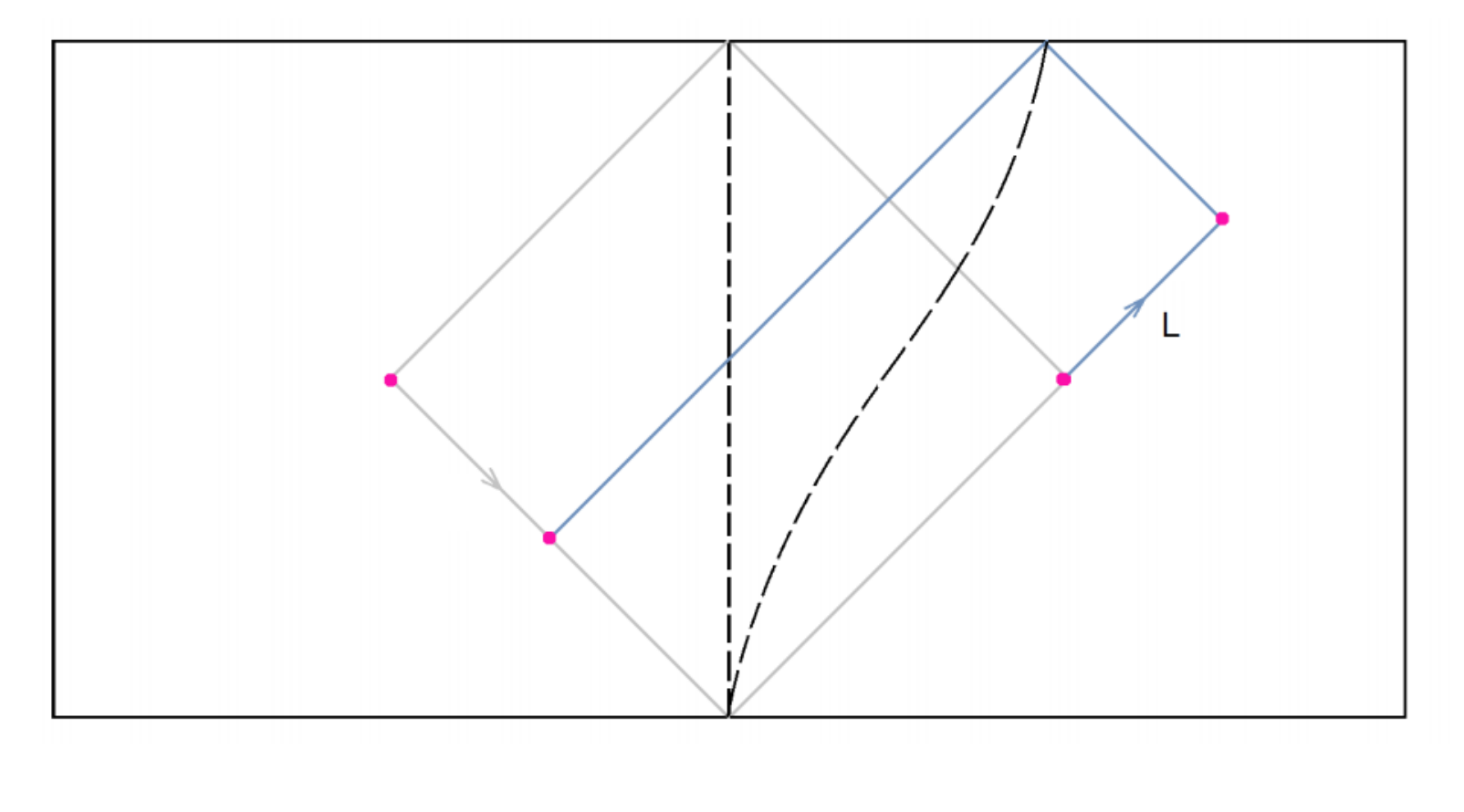}
\caption{Two-dimensional \dS with two static patches related by a light-like generator $L.$ }
\label{lightshift}
\end{center}
\end{figure}

The group $O(d,1)$ relates different static patches within a single de Sitter space. For example in fig \ref{lightshift} shows two static patches related by the action of a light-like generator $L.$ 

\bn
If the $O(d,1)$ symmetry were not broken by the GKS anomaly one would expect that the dynamics in the two patches would be identical. In particular the Hamiltonian in one patch would be related to that in the other patch by a unitary transformation and the two spectra would be identical. However our result suggests that this is not true; what is more likely is that the coarse-grained spectra of the two Hamiltonians are the same but at the discrete level of individual eigenvalues, the spectra do not match. The occurrence and timing of fluctuations in the two static patches (for example quantum recurrences)  would be different. The Hamiltonians for different patches might be drawn from a single ensemble but would be different instances of that ensemble. In the absence of knowledge about which Hamiltonian governs an observer's patch, averaging over the ensemble would make sense\footnote{Similar ideas have long been advocated by Banks and Fischler  \cite{Banks:2002wr}\cite{Banks:2003cg} who argues that many static patch  Hamiltonians may lead to identical observations. Their argument is quite different from the one in this paper. It is based on assumed limits of observation in a closed world with finite entropy.}.

\subsection{Gauge Symmetry?}

It is usually assumed that the $O(d,1)$  relating different static patches is a  gauge symmetry, or a redundancy of the description. But having a gauge symmetry requires that the gauge transformation can be  consistently defined. This does not appear to be possible  for the symmetries of de Sitter space except in the
semiclassical limit.  The GKS anomaly precludes the existence of the generators to higher order in $e^{-S_0}$ and indicate the different static patches are  inequivalent when effects of order $e^{-S_0}$ are considered. Thus, the answer to the question is that $O(d,1)$ should be treated as a gauge symmetry in the semiclassical approximation, but not in the full non-perturbative theory. One could select a  static patch not by gauge fixing, but by simply selecting the patch whose detailed energy levels have some specific pattern.

One possible conclusion is that eternal  de Sitter space is not consistent---a view  taken in \cite{Goheer:2002vf}.
In this paper I've advocated another viewpoint; namely  that the de Sitter symmetries are  approximate, valid in classical theory and in perturbation theory, but not beyond. In fact for other topologies the action of $O(d,1)$ may not even be defined. For example it is hard to imagine  the action of
$O(d,1)$
on the Nariai space $S(2) \times S(2).$

In some ways
the situation seems similar to recent discussions of global symmetries and their breaking by higher topologies  \cite{Chen:2020ojn} where it was suggested that  global symmetries (forbidden by gravity)  may be restored in the ensemble average. In the context of \dS the average over all the static patches in  \dS might have the symmetry that the individual instances don't have. 

\section{Conclusion}
Eternal de Sitter space is a spacetime without time-like boundaries, but a static patch is bounded by a horizon. On the basis of covariant entropy bounds I argued that the natural place to locate holographic \dof is on the stretched horizon. The only things that happen in eternal de Sitter space are fluctuations of these horizon degrees of freedom,  which from the bulk point of view sporadically produce interesting objects deep in the interior of the static patch.

We've explored three non-perturbative de Sitter space phenomena, related to these fluctuations. The first:  the violation of the de Sitter symmetry $O(4,1)$ due to the GKS anomaly.  The variance $\rm Var \it(L_-)$  in \ref{variance} is a  measure of the magnitude of the violation.

The second: large scale Boltzmann fluctuations in which the holographic horizon degrees of freedom undergo freak rearrangements, leading to  large black holes materializing in the interior of the static patch. The probability for this to happen is given by \ref{prob}. The second term of this expression is non-perturbative and represents the creation of the largest  black holes with entropy $\approx S_0/3.$

Finally wormholes and higher genus geometries:    a saddle point  due to the   Nariai geometry $S_2\times S_2$  contributes non-perturbatively to the gravitational path integral and describe a massive fluctuation in which  de Sitter space turns itself ``inside out."

These phenomena which all scale exponentially with $-S_0$   are closely connected. One might even say they  express  the same 
 underlying fact; namely the discreteness of the energy spectrum, and finite level spacing $\delta E \sim e^{-S}.$ Moreover they are extensions of things that have been observed in other contexts such as SYK and JT gravity.
The new thing here is that they appear in the holography of de Sitter space.

The violation of symmetry is especially interesting. It means that different static patches are inequivalent. They may have different Hamiltonians and different energy spectra although the coarse grained spectra must be the same to insure a universal semiclassical limit.  Although it is an open question, it is an interesting conjecture that the symmetry-violation is washed out by some form of ensemble averaging, as is thought  to be the case for JT gravity.

\section*{Acknowledgements}

I am grateful to  
Adel  Rahman, and Douglas Stanford for  discussions, and especially to  Adam Brown for discussions about the Nariai saddle point.

\end{document}